\documentclass[article,twocolumn,floats,nofootinbib,nobibnotes,superscriptaddress,10pt]{revtex4-1}

\usepackage{graphicx,amsmath,amssymb,hyperref}
\hypersetup{colorlinks=true, linkcolor=red, citecolor=magenta, urlcolor=blue}
\usepackage{dcolumn}
\usepackage{xcolor,colortbl}
\usepackage{enumitem}
\usepackage{eqnarray}
\usepackage{makecell}
\usepackage{cancel}
\usepackage{etoolbox}
\usepackage{comment}
\usepackage{amsfonts,mathtools}
\usepackage{tabularx,booktabs}
\usepackage{multirow}
\usepackage{rotating}
\usepackage{array}
\newcolumntype{P}[1]{>{\centering\arraybackslash}p{#1}}
\usepackage{makecell}
\usepackage{refcount}
\usepackage{makecell}

\usepackage{ulem} 
\usepackage{tikz}
\usepackage{bbding}
\usepackage{cancel}

\newcommand{\notcheckmark}{{$\surd$}\textsuperscript{\textcolor{black}{\kern-0.35em{\bf--}}}}

\usepackage{pifont}
\newcommand{\cmark}{\ding{51}}  
\newcommand{\xmark}{\ding{55}}  

\makeatletter
\newcommand\footnoteref[1]{\protected@xdef\@thefnmark{\ref{#1}}\@footnotemark}
\makeatother

\begin{document}

\title{\texttt{oLIMpus}: An Effective Model for Line Intensity Mapping\\
Auto- and Cross-Power Spectra in Cosmic Dawn and Reionization
}

\newcommand\affSL{  
\affiliation{Department of Physics, Ben-Gurion University of the Negev, Be'er Sheva 84105, Israel}
}

\newcommand\affEK{
\affiliation{Department of Physics, Ben-Gurion University of the Negev, Be'er Sheva 84105, Israel}
\affiliation{Texas Center for Cosmology and Astroparticle Physics, Weinberg Institute, Department of Physics,
The University of Texas at Austin, Austin, TX 78712, USA}
}

\newcommand\affJM{
\affiliation{Department of Astronomy, The University of Texas at Austin,
2515 Speedway, Stop C1400, Austin, Texas 78712, USA}}

\author{Sarah Libanore}
\email{libanore@bgu.ac.il}
\affSL

\author{Julian B. Mu\~noz}
\affJM

\author{Ely D. Kovetz}
\affEK

\begin{abstract}
Line-intensity mapping (LIM) is emerging as a powerful probe of the high-redshift Universe, with a 
growing number of LIM experiments targeting various spectral lines  
deep into the epochs of reionization and cosmic dawn.  
A key remaining challenge is the consistent and efficient modeling of the diverse emission lines and of the observables of different surveys. 
Here, we present \texttt{oLIMpus}\footnote{\label{olimpuslink}\href{https://github.com/slibanore/oLIMpus}{\includegraphics[height=1em]{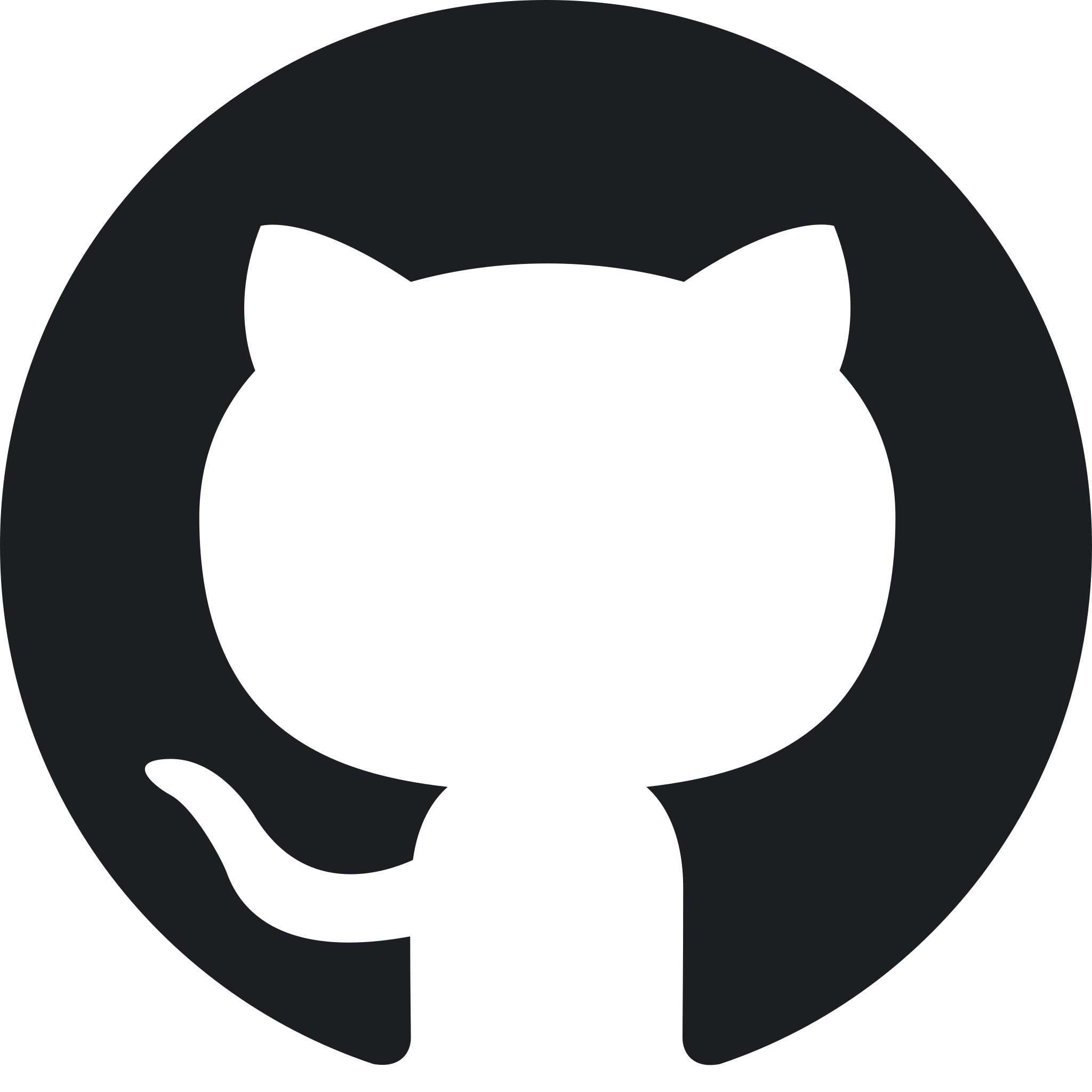} \texttt{https://github.com/slibanore/oLIMpus}}}, a fully analytical effective model to study LIM auto- and cross- power spectra. Our work builds on the 21-cm effective model presented in \texttt{Zeus21}, applying it to star-forming lines and improving it in different aspects. Our code accounts for shot noise and linear redshift-space distortions and it includes by default prescriptions for OII, OIII, H$\alpha$, H$\beta$, CII, CO line luminosities, together with the 21-cm model inherited from \texttt{Zeus21}. Beyond auto- and cross-power spectra, \texttt{oLIMpus} can produce mock coeval boxes and lightcones,  
and with a computational time of $\sim s$ 
it is ideal for parameter-space exploration and inference. 
Its modular implementation makes it easy to customize and extend, enabling various applications, such as MCMC analyses and consistent multi-line cross-correlations. 
\end{abstract}

\maketitle

\section{Introduction}\label{sec:intro}

Cosmic dawn and the epoch of reionization (EoR) represent crucial periods in the history of the Universe, holding the key to understanding the interplay between astrophysical and cosmological processes shaping the young Universe, including the formation of the first luminous structures. Line-intensity mapping (LIM, see e.g.,~reviews in Refs.~\cite{kovetz2017lineintensitymapping2017status,Bernal:2022jap}) offers a powerful observational strategy to probe these epochs, capturing the cumulative emission from the entire galaxy population, including unresolved sources, thus providing access to the faint end of their luminosity functions, e.g.,~Refs.~\cite{Suginohara:1998ti,Visbal:2010rz,Breysse:2015saa}, the underlying cosmological parameters, e.g.,~Refs.~\cite{Fonseca:2016qqw,Schaan:2021hhy,Chang:2007xk,Pritchard:2011xb,Bernal:2019gfq} and unlocking small-scale information that is otherwise difficult to probe, e.g.,~Refs.~\cite{Libanore:2022ntl,Adi:2023qdf}. As a rapidly growing number of LIM experiments are coming online, we are approaching an era where data from these early epochs will become increasingly accessible. 

In the near future, LIM experiments will gain access to signals from a broad range of spectral lines spanning an extensive redshift interval. Among the most exciting recent developments is the launch of the Spectro-Photometer for the History of the Universe, Epoch of Reionization, and Ices Explorer (SPHEREx,~\cite{SPHEREx:2014bgr}), which is observing rest-frame UV and optical lines (e.g.,~OII, OIII and the Balmer lines H$\alpha$ and H$\beta$) up to redshift $z\sim 9$. SPHEREx also observes Lyman-$\alpha$ at even higher redshifts. At intermediate redshifts ($z\sim 2-4$), Ly$\alpha$ has already been observed by the Hobby-Eberly Telescope Dark Energy Experiment  (HETDEX~\cite{Hill:2008mv,Gebhardt:2021vfo}), with ULTRASAT~\cite{benami2023scientificpayloadultraviolettransient,shvartzvald2023ultrasatwidefieldtimedomainuv,Libanore:2024wvv} soon joining to cover even lower redshifts ($z<2$). 
On the far-infrared side, the CII line was recently targeted up to $z>6$ by CONCERTO~\cite{CONCERTO2020,Bethermin:2022lmd}, while the Tomographic Ionized-carbon Mapping Experiment (TIME,~\cite{Sun:2020mco,TIME}) and the Fred Young Submillimeter Telescope (FYST~\cite{Karoumpis_2022}) are expected to follow, focusing on $z\sim 6-9$ and $z\sim 4-8$, respectively. 
Lower-frequency observations of the CO rotational lines are also underway. The Carbon Monoxide Mapping Array Project (COMAP,~\cite{COMAP:2021nrp}) is currently observing the CO(1–0) line at $z\sim 3$ as well as fainter contributions of CO(2–1) from higher redshifts. The collaboration plans to add a lower frequency detector to ultimately access these two transitions from the full $z\sim 5-8$ range~\cite{COMAP:2021nrp}. 
Other spectral lines, which originate from different phases of the interstellar medium and various stages of star formation, such as He II~\cite{Visbal:2015sca,Parsons:2021qyw,Serra:2016jzs} or NII~\cite{Padilla:2022asq}, although less commonly discussed in the literature, may also fall within the observational windows of  upcoming instruments.

In parallel, 21-cm cosmology~\cite{Pritchard:2011xb,Barkana:2000fd,Furlanetto:2006jb} is advancing rapidly, particularly thanks to the precursors of the Square Kilometer Array Observatory (SKAO,~\cite{Braun:2015zta,Braun:2019gdo}). Interferometers such as the Hydrogen Epoch of Reionization Array (HERA~\cite{DeBoer:2016tnn}) and the Low-Frequency Array (LOFAR~\cite{LOFAR:2013jil}) are targeting the 21-cm power spectrum during the EoR and cosmic dawn. At these early times, the 21-cm signal is sourced in the intergalactic medium (IGM), and its evolution is driven by the radiation from the first galaxies, depending on the way it heats the gas and its efficiency in ionizing the IGM~\cite{Furlanetto:2004nh,Morales:2009gs}. Upper limits on the 21-cm power spectrum have already been established by HERA~\cite{HERA:2021bsv,HERA:2021noe,Lazare:2023jkg} and LOFAR~\cite{Ceccotti:2025bcd,Ghara:2025xzu}, {as well as by other interferometers, i.e.,~the Murchison Widefield Array (MWA}~\cite{EwallWice_2016,Yoshiura_2021}{), the Owens Valley Long Wavelength Array (LWA}~\cite{Eastwood:2019rwh,Garsden_2021}{), and the New Extension in Nan\c{c}ay Upgrading LOFAR (NenuFAR,}~\cite{Munshi:2023buw,Munshi:2025hgk}{)}. More stringent constraints are expected in the near future.

In this context, combining multiple spectral lines offers a powerful strategy to extract robust information. This multi-tracer approach is emerging as a promising avenue to leveraging the full potential of the rapidly growing LIM observational landscape,~see e.g.,~Refs.~\cite{Serra:2016jzs,Kannan:2021ucy,Moriwaki:2019dbg,Moriwaki:2024kvp,Mas-Ribas:2022jok,Sun:2022ucx,Sun:2024vhy,Padmanabhan:2021tjr,Chang:2019xgc,Fronenberg_2024}. 

However, interpreting LIM signals poses significant challenges, both from the theoretical and observational points of view. Particularly relevant is the question of how to consistently model the emission of different lines. LIM simulations present an intrinsic multi-scale problem, since they need to connect processes that take place on subgalactic scales with large-scale structure building up across cosmological volumes. Efforts in this respect have been made with varying levels of complexity and accuracy, ranging from full radiative-transfer hydrodynamical simulations,~e.g.~Refs.~\cite{THESAN_1,THESAN_2,THESAN_3,Lee:2025vwq}, to semi-analytic simulations such as \texttt{21cmFAST}~\cite{Mesinger_2011,Murray:2020trn} and \texttt{LIMFAST}~\cite{Mas-Ribas:2022jok,Sun:2022ucx,Sun:2022ucx}, and to scaling relations calibrated on simulations or data, e.g.~Refs.~\cite{yang2025newframeworkismemission,Lagache_2018,Li:2015gqa,Yang:2021myt,Sun_2019}, which are used to either paint halos with line luminosities, e.g.,~Refs.~\cite{Sato-Polito:2022wiq,Yang:2020lpg}, or to inform analytical models, e.g.,~Refs.~\cite{Zhang:2023oem,Bernal:2019jdo}. 

Recently, Refs.~\cite{Munoz:2023kkg,Cruz:2024fsv} introduced \texttt{Zeus21}\footnote{\url{https://github.com/JulianBMunoz/Zeus21}}, an effective model to fully analytically compute the 21-cm global signal and fluctuations in seconds. {\texttt{Zeus21} was originally developed to compute the 21-cm power spectrum during cosmic dawn, but it is currently being extended to the Epoch of Reionization~\cite{Sklansky:2025}. The code }
accounts for non-locality and non-linearity in the star formation rate density by approximating this quantity by a lognormal random field (e.g.\ Ref.~\cite{Coles_1991}).
Although this approach inevitably introduces some approximations, it has been shown that \texttt{Zeus21} results are in agreement with semi-analytic simulations such as \texttt{21cmFAST}, while being orders of magnitude faster and computationally cheaper. 

\begin{figure*}[ht!]
   \includegraphics[width=\linewidth]{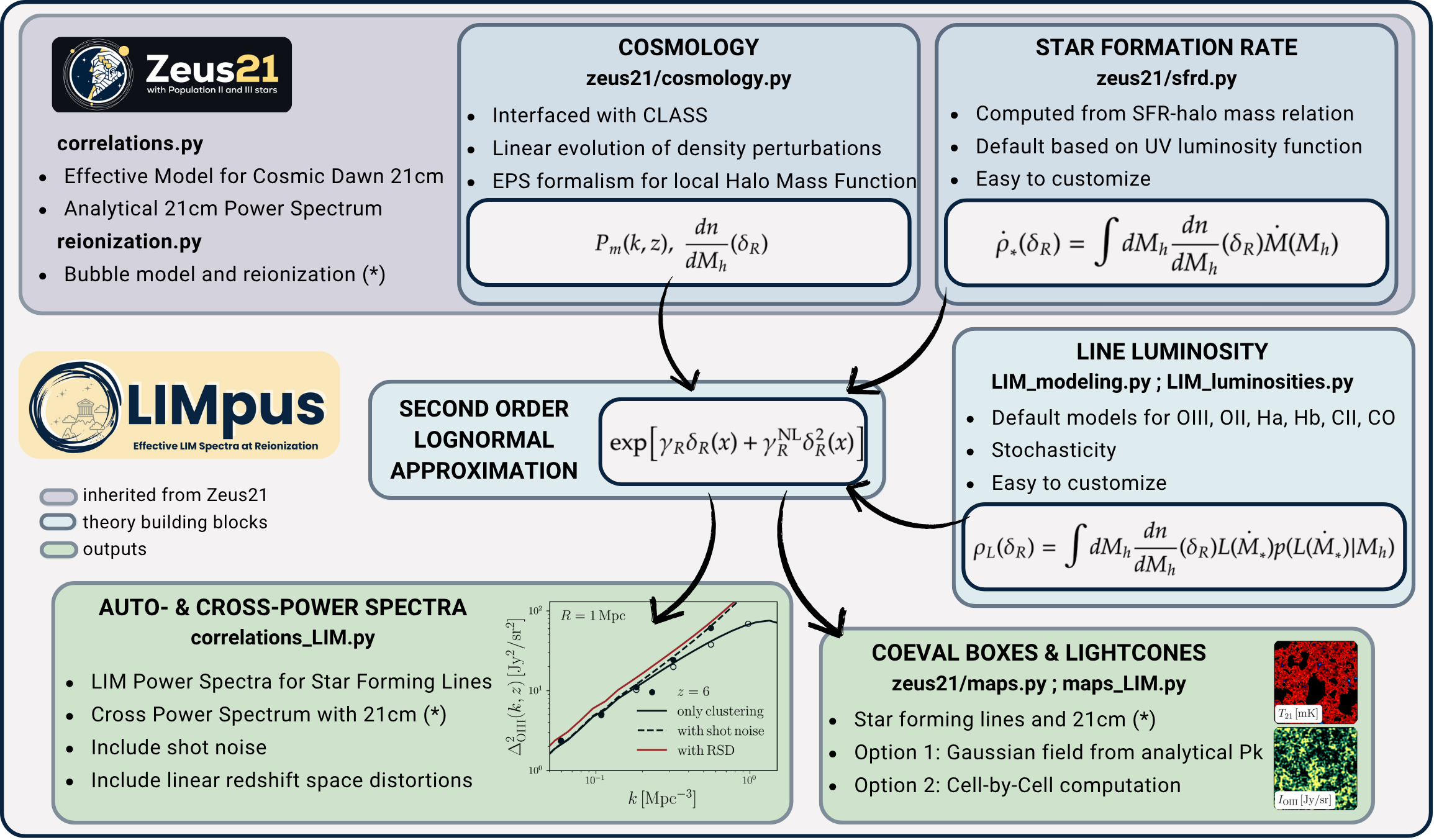}
   \caption{Flowchart of \texttt{oLIMpus}. The top panel indicates the building blocks inherited from \texttt{Zeus21} to model the {\bf cosmology} and the {\bf star formation rate} $\dot{M_*}(M_h)$.
   On top of these, \texttt{oLIMpus} introduces a new building block to model the {\bf line luminosity} of the star forming lines of interest $L(\dot{M}_*(M_h))$, for which we also account for lognormal scattering. Combined with $dn/dM_h$, the star formation rate and the line luminosity provide the star formation rate density $\dot{\rho}_{*}$ and the luminosity density $\rho_L$. The central panel recalls the key assumption in \texttt{oLIMpus}: both $\dot{\rho}_{*}$ and $\rho_L$ can be estimated as {\bf lognormal functions of the smoothed density} field $\delta_R$. The approximation $\dot{\rho}_{*}(\delta_R)\propto e^{\gamma_R\delta_R}$ was already introduced in \texttt{Zeus21} to estimate the evolution of the 21-cm signal; here, we improve over the previous work by adding the second-order dependence on $\delta_R^2$, and by extending this formalism to the star-forming lines. Moreover, \texttt{Zeus21} can now be used as a submodel of \texttt{oLIMpus}, hence producing the 21-cm signal during cosmic dawn and the epoch of reionization consistently with the evolution of the other lines. Finally, the green panels in the bottom row represent the main outputs of our code: the lognormal approximation allows us to estimate fully analytically the {\bf auto- and cross- power spectra} of star forming lines, including shot noise and redshift space distortions. The auto-power spectra, in turn, can be used to produce {\bf coeval boxes and lightcones} of the expected LIM signal, which correlate with the underlying density field. \texttt{oLIMpus} can also produce coeval boxes and lightcones through a cell-by-cell computation of the relevant quantities, including the reionization field that drives the evolution of the ionizing bubble and their effect on the 21-cm signal. The reionization model is part of an upcoming new release of \texttt{Zeus21}~\cite{Sklansky:2025}; for the moment, \texttt{oLIMpus} relies on a reduced version, which enters in the steps indicated with (*) in the flowchart; these will be updated in subsequent work. }
   \label{fig:code_summary}
\end{figure*}

In this work, we extend the \texttt{Zeus21} formalism and  apply it to model the LIM power spectra of lines whose emission is associated with star formation processes. In our case, the building block is the line luminosity density, which we model using a second-order lognormal function of the density. We implement and release our approach in the publicly available code \texttt{oLIMpus}\footnoteref{olimpuslink}. 
The main advantages of \texttt{oLIMpus} are the following:
\begin{itemize}[itemsep=1pt,topsep=5pt]
    \item It provides a consistent framework to model line intensities during cosmic dawn and the epoch of reionization, including both star-forming lines (OIII, OII, H$\alpha$, H$\beta$, CII, CO) and 21-cm.
    \item Its analytical formalism makes it possible to produce LIM non-linear auto- and cross-power spectra in seconds, through a computationally efficient algorithm that makes 
    it the perfect tool to explore different models and large parameter spaces, for example in the context of MCMC or other Bayesian inference analyses. 
    \item In this first release, line luminosities are modeled based on scaling relations calibrated on state-of-the-art simulations. The lean formalism at the core of the code makes it very easy to customize these prescriptions, potentially expanding them to include more refined line models. 
\end{itemize}
Figure~\ref{fig:code_summary} summarizes the rationale behind \texttt{oLIMpus}, highlighting the main ingredients our code relies on. 

The rest of the paper is structured as follows. Section~\ref{sec:nutshell} outlines the main structure of \texttt{oLIMpus}, providing a qualitative description that can be used as a guideline throughout the rest of the text. In Sect.~\ref{sec:line_model} we introduce our formalism, which in Sect.~\ref{sec:2pt} we apply to model auto- and cross- power spectra of different lines, including 
its application for parameter variations in Sect.~\ref{sec:science}. In Sect.~\ref{sec:compare_codes} we compare our results 
with other publicly available codes  
to produce LIM power spectra, and in Sect.~\ref{sec:maps} we describe how to produce mock coeval boxes and lightcones of relevant quantities. Finally, in Sect.~\ref{sec:conclusion}, we comment on other possible applications and science cases for which \texttt{oLIMpus} seems to be particularly well suited.  
Throughout the text, we rely on the {\it Planck 2018} cosmological parameters. 
We indicate with $\rho_{m,0}=\rho_c\Omega_m=2.78\times 10^{11}h^2\Omega_c\,M_\odot/{\rm Mpc}^3$ the matter density today, and with $\delta_c=1.686$ the critical density barrier for collapse.


\section{\texttt{oLIMpus} in a nutshell}\label{sec:nutshell}

We begin by providing an overall description of our methodology and formalism, with the aim of helping the reader navigate the rest of the paper. Detailed definitions and derivations of all the quantities we mention here are provided in the subsequent sections. 

The \texttt{oLIMpus} code  builds on the pre-existing \texttt{Zeus21} code~\cite{Munoz:2023kkg,Cruz:2024fsv}, improving its formalism and extending it to the context of LIM science. 
Through our new pipeline, which now includes \texttt{Zeus21} as a submodule, we can produce the following main outputs:
\begin{enumerate}[itemsep=1pt,topsep=5pt]
    \item The analytical non-linear\footnote{``Non-linear" refers to the power spectrum dependence on the star formation rate density and line luminosity density. Instead, cosmological density perturbations are evolved in linear theory, consistently with the extended Press Schechter theory that requires the linear $\delta$ in the computation of Eq.~\eqref{eq:corr_EPS}.} auto- and cross-power spectra of star-forming lines in $\mathcal{O}(\rm seconds)$ runtime.
    \item Coeval boxes and lightcones of the density field, the star formation rate density (SFRD), the intensity of star-forming lines, all 
    based on analytical effective models. All of these have coherent statistical and astrophysical properties, hence can be consistently used for cross-correlation studies. These boxes can be produced in $\mathcal{O}(5\,{\rm minutes})$ for an array of 100 redshift values, with $L_{\rm box}=300\,{\rm Mpc},\,N_{\rm cell}=150$ (i.e.\ $\sim2\,{\rm Mpc}$ resolution), easily varying the underlying cosmological, astrophysical and line models.
    \item Coeval boxes and lightcones 
    based on a cell-by-cell computation 
    of the SFRD and of the line luminosity density; this (slower) approach allows us to produce mock-simulations against which our effective model 
    can be tested. 
\end{enumerate}
Moreover, \texttt{oLIMpus} can output the 21-cm power spectrum at cosmic dawn computed using \texttt{Zeus21}, and use it to produce coeval boxes and lightcones of the 21-cm brightness temperature. {A new version of \texttt{Zeus21}~\cite{Sklansky:2025} extends the computation of the 21-cm signal into the EoR; currently, a preliminary version allows the user to generate maps of the reionization field, by computing the neutral fraction cell-by-cell. Correlations can then be estimated on the map level; a fully analytical treatment of the 21-cm EoR auto-power spectrum and of its cross correlation with other lines is currently under development. }
Being able to consistently model the star-forming and 21-cm lines in overlapping redshifts opens the window to cross-correlation studies; we will widely explore them in our upcoming work~\cite{Libanore:2025gte}.  

\subsection{Summary of the effective model}

The formalism of \texttt{oLIMpus} builds upon the one in \texttt{Zeus21}. 
The evolution of the 21-cm signal during cosmic dawn is driven by non-local (both spatially and temporally) integrals of the star formation rate density (SFRD, $\dot{\rho}_*$); therefore, it can be estimated in terms of their weighted sums over different smoothing radii $R$ and redshifts $z$. Since the SFRD traces the process of structure formation, which is known to be well described by a lognormal random field~\cite{Coles_1991}, it  can be effectively modeled as a lognormal function of the density when smoothed over $R$~\cite{Munoz:2023kkg},
$
    \left<\dot{\rho}_{*}\right>_R\propto e^{\gamma_R\delta_R},
$
where $\gamma_R$ plays the role of an effective non-linear bias, see Sect.~\ref{sec:2pt}. The statistical properties of the density field $\delta$ can be (at first approximation) modeled using the linear matter power spectrum, $P(k,z)=D^2(z)P_m(k,z=0)$, where $D(z)$ is the growth factor, thus depending on the underlying cosmological model. Both $D(z)$ and $P_m(k,z=0)$ are obtained using the Boltzmann solver \texttt{CLASS}\footnote{\url{https://github.com/lesgourg/class_publi}c}~\cite{Blas:2011rf}. 
Moreover, the extended Press-Schechter formalism~\cite{Press:1973iz,Bond:1990iw} allows one to link the SFRD to the density-modulated halo mass function (HMF, $dn/dM_h(\delta_R)$).

As we will detail in Sect.~\ref{sec:2pt}, the lognormal approximation of the SFRD provides an easy way to analytically estimate its two-point function, and consequently the two-point functions of the quantities that depend on it. These can finally be combined in order to estimate the 21-cm two-point function and its power spectrum.

In this work, we extend this formalism to model the power spectra of other emission lines, which are associated with star formation. The intensity of these lines as a function of redshift  ($I_\nu(z)$) can be estimated based on their luminosity density ($\rho_L$, see details in Sect.~\ref{sec:line_model}), which we aim to effectively model in analogy to the SFRD. 
To do so, we need to account for some subtleties:
\begin{itemize}[itemsep=1pt,topsep=5pt]
    \item The line luminosity can be modeled as a non-linear function of the star formation rate and of the host halo mass, $L(\dot{M}_*(M_h),M_h)$. To account for such non-linearities, we apply the lognormal approximation directly to the luminosity density, $\rho_L\propto e^{\gamma_{L}\delta_R}$, instead of estimating it from the SFRD.
    \item The emission of star-forming lines is a local process, which does not require summing over different radii $R$. In our case, therefore, for each line we consider a single smoothing radius $R_0$, which represents either the scale below which the lognormal approximation breaks, or the resolution of the LIM observation (when the user chooses to account for it explicitly).
    \item Because LIM traces smaller physical scales and lower redshifts than 21-cm, matter overdensities $\delta$ can be larger. For this reason, we extend the lognormal approximation to second order,
    \begin{equation}
        \langle\rho_L\rangle \propto e^{\gamma_{R}\delta_R+\gamma_{R}^{\rm NL}\delta_R^2},
    \end{equation}
    and derive its two-point function accordingly (note that the value of $\gamma_R$ in the $\rho_L$ case is different from its value in the SFRD case).
    We  introduce a similar extension to the SFRD calculation, to improve its modeling during the end stages of reionization. 
    \item Since star-forming lines are mainly emitted inside galaxies, namely from discrete sources, and not smoothed over large radii, the contribution of the shot noise (i.e.,~Poisson noise) to their small-scale power spectra cannot be neglected.
\end{itemize}

Accounting for all the previous points, we built an effective model for the power spectrum of star-forming lines and we implemented it into our code \texttt{oLIMpus} (see Fig.~\ref{fig:code_summary} for a visual summary of the implementation). 
We now proceed with a more detailed description of our formalism, carefully testing the reliability of \texttt{oLIMpus} through comparisons with other LIM simulations and analytical codes available in the literature.


\section{Line luminosity density}\label{sec:line_model}

The primary building block of \texttt{oLIMpus} is the computation of the line  luminosity, $\rho_L\,[L_\odot/{\rm Mpc^3}]$. This shares a similar structure with the SFRD $\dot{\rho}_*\,[M_\odot/{\rm yr/Mpc^3}]$ building block of \texttt{Zeus21}; for this reason, in the following we generalize the derivation using the generic quantity $\rho$, which interchangeably represents $\rho_L$ and $\dot{\rho}_*$.

The quantity $\rho$ is estimated as a function of the mass of the host dark matter halo through the function $f(M_h,z)$, namely either the line luminosity $L(M_h)\,[L_\odot]$ or the star-formation rate $\dot{M}_*(M_h)\,[M_\odot/\rm yr]$. All the plots in the paper, except when differently specified, are produced using $f(M_h,z)=L_{\rm OIII}(M_h,z)$, where the OIII luminosity is from Ref.~\cite{yang2025newframeworkismemission}, see details in Appendix~\ref{app:lines}.

The average value of $\rho$ is computed as 
\begin{equation}\label{eq:rho_Lag}      
    \langle{\rho}^{\rm Lag}(z)\rangle=\int dM_h(z) \frac{dn}{dM_h}f(M_h,z),
\end{equation}
where $dn/dM_h$ is the halo mass function (HMF). We rely on the Sheth-Mo-Tormen formalism~\cite{Sheth:1999su}, in which 
\begin{equation}\label{eq:HMF}
    \frac{dn}{dM_h}(z)=-A_{\rm ST}\sqrt{\frac{2}{\pi}}\frac{\rho_{m,0}}{\sigma_{M_h} M_h}\frac{d\sigma_{M_h}}{dM_h}\nu(1+\nu^{-2p_{\rm ST}})e^{-\nu^2/2},
\end{equation}
where $\nu=\sqrt{a_{\rm ST}}\,\delta_c/\sigma_{M_h}$, while $\sigma_{M_h}^2(z)$ is the variance of the matter fluctuations on scale $R_h = (3M_h/4\pi)^{1/3}$, and the parameters $a_{\rm ST}=0.707,\,A_{\rm ST}=0.3222,\,p_{\rm ST}=0.3$ are calibrated by fitting high-$z$ N-body  simulations~\cite{Schneider:2018xba}. Note that because of  the way the HMF is calibrated, $\langle{\rho}^{\rm Lag}(z)\rangle$ in Eq.~\eqref{eq:rho_Lag} is the average in Lagrangian space. 

To develop our statistical analysis, on one side we need to connect $\rho(z)$ to the local matter fluctuations $\delta(z)=D(z)\delta_{m,0}$, eventually smoothed over regions of comoving radius $R$, which we label as $\delta_R(z)$. On the other side, we want to obtain $\rho(z)$ in Eulerian space. Following Refs.~\cite{Barkana_2005,Mesinger_2011}, we accomplish this result by defining:
\begin{equation}\label{eq:rho_deltaR}
\begin{aligned}
        \rho(z|\delta_R)&=(1+\delta_R)\rho^{\rm Lag}(z|\delta_R)\\
    &=(1+\delta_R)\int dM_h\frac{dn}{dM_h}\mathcal{C}_{\rm EPS}f(M_h,z),
\end{aligned}
\end{equation}
where two correction factors are introduced. 

\noindent
First of all, in the quantity 
\begin{equation}\label{eq:corr_EPS}
    \mathcal{C}_{\rm EPS}=\frac{dn}{dM_h}\frac{dn^{\rm EPS}/dM_h}{\langle dn^{\rm EPS}/dM_h\rangle}
\end{equation}
we make use of the HMF in the EPS formalism~\cite{Press:1973iz,Bond:1990iw}, $dn^{\rm EPS}/dM_h$, and its average value $\langle dn^{\rm EPS}/dM_h\rangle$. These can be related to $\delta_R$ and its variance $\sigma_R^2$ through~\cite{Lacey_1994}
\begin{align}
    \frac{dn^{\rm EPS}/dM_h}{\langle dn^{\rm EPS}/dM_h\rangle} &= \frac{\tilde{\nu}}{{\nu}_0}\frac{\sigma_R^2}{\tilde{\sigma}^2}e^{a_{\rm ST}(\tilde{\nu}^2-\nu^2_0)/2},\\
    \tilde{\nu} = \frac{\delta_c-\delta_R}{\tilde{\sigma}},\quad \nu_0 &=\frac{\delta_c}{\sigma_{M_h}},\quad \tilde{\sigma}=\sqrt{\sigma_{M_h}^2-\sigma_R^2}.
\end{align}
Rescaling this factor by the Sheth-Mo-Tormen HMF $dn/dM_h$ in Eq.~\eqref{eq:corr_EPS} allows us to recover the correct average value when integrating over $dM_h$.

Secondly, in Eq.~\eqref{eq:rho_deltaR} the factor $(1+\delta_R)$ transforms $\rho$ from Lagrangian to Eulerian space. When performing this transformation, however, we need to be careful since the average value gets affected as well. To solve this issue, we define the average in Eulerian space $\bar{\rho}(z)$, as
\begin{equation}\label{eq:phi_LtoE}
    \bar{\rho}(z) = \frac{\langle(1+\delta_R)\rho^{\rm Lag}(z|\delta_R)\rangle}{\langle\rho^{\rm Lag}(z)\rangle}\langle\rho^{\rm Lag}(z)\rangle = \phi_{\rm LtoE}\langle\rho^{\rm Lag}(z)\rangle,
\end{equation}
where $\phi_{\rm LtoE}$ indicates the correction between the averages in the two spaces (see e.g., Appendix A in Ref.~\cite{Munoz:2023kkg}). 

Equation~\eqref{eq:rho_deltaR} already provides a tool to produce mock luminosity density (or SFRD) coeval boxes. Starting from a density coeval box $\delta(\vec{x},z)$ (see Sect.~\ref{sec:maps}) and applying the $\delta_R-\rho$ relation in each cell, we can indeed produce a coeval box of luminosity density (or SFRD) fluctuations. To recover the expected average, first we produce a box in Lagrangian space, i.e.,~we apply Eq.~\eqref{eq:rho_deltaR} without the $(1+\delta_R)$ factor to obtain $\rho^{\rm Lag}(\vec{x},z)$, 
Then, we rescale the box by $(1+\delta(\vec{x},z))$ to obtain the Eulerian $\rho(\vec{x},z)$; the $\phi_{\rm LtoE}$ correction from Eq.~\eqref{eq:phi_LtoE} is implicitly accounted for in this procedure. Finally, we smooth the box over $R$ by applying a top hat filter in real space.

Even if faster than a full simulation, this approach still lacks the computational efficiency we are looking for, particularly when producing large boxes with high resolution. In the following, therefore, we proceed by introducing the lognormal approximation for $\rho$ that will allow us to estimate the analytical two-point function. From now on, we will use the boxes produced through the cell-by-cell algorithm as a benchmark to test the goodness of our approximations and the self-consistency of the code.


\subsection{Lognormal approximation}\label{sec:lognormal}

Following a similar reasoning as in \texttt{Zeus21}, we now want to approximate the Eulerian $\rho(z|\delta_R)$ as a lognormal function of the Gaussian random variable $\delta_R(z)$. Ref.~\cite{Munoz:2023kkg} already showed that $\dot{\rho}_*(z|\delta_R)\propto e^{\gamma_R\delta_R}$ captures well the SFRD dependence on $\delta_R$ during the cosmic dawn and on scales $R\gtrsim 3\,{\rm Mpc}$, where $\delta_R(z)\ll\delta_c$. To apply a similar effective model to the luminosity density required in LIM studies, however, we need to go beyond this approximation, extending it to smaller $R$ and lower $z$, where $\delta_R(z)$ becomes larger. Thus, we rely on a second-order lognormal expansion, where
\begin{equation}\label{eq:lognormal}
    \rho(z|\delta_R) =\bar{\rho}(z) \frac{\exp[\gamma_R(z)\delta_R(z)+\gamma_R^{\rm NL}(z)\delta^2_R(z)]}{\mathcal{N}(z)}.
\end{equation}
By taking advantage of the expression for the derivative of the exponential, we can compute the Eulerian $\gamma_R(z)$ and $\gamma_R^{\rm NL}(z)$ coefficients straightforwardly as  
\begin{equation}\label{eq:gammas}
\begin{aligned}
    \gamma_R(z) = \frac{d\log\rho(z|\delta_R)}{d\delta_R}&\biggl|_{\delta_R=0}>0,\\ 
    \gamma_R^{\rm NL}(z) = \frac{d^2\log\rho(z|\delta_R)}{d\delta_R^2}&\biggl|_{\delta_R=0}<0.
\end{aligned}
\end{equation}
The factor $\mathcal{N}(z)$ in Eq.~\eqref{eq:lognormal} normalizes the distribution,
\begin{equation}
\frac{\langle\exp[\gamma_R(z)\delta_R(z)+\gamma_R^{\rm NL}(z)\delta^2_R(z)]\rangle}{\mathcal{N}(z)}=1,
\end{equation}
so it is defined as
\begin{equation}
\begin{aligned} 
\mathcal{N}(z)&=\int d\delta_R \frac{\exp\left[\gamma_R\delta_R+\gamma_R^{\rm NL}\delta_R^2\right]}{{\sqrt{2\pi}\sigma_R}}\exp\left(\frac{-\delta_R^2}{2\sigma_R^2}\right)\\
&=\frac{1}{\sqrt{1-2\gamma_R^{\rm NL}\sigma_R^2}}\exp\left(\frac{\gamma_R^2\sigma_R^2}{2-4\gamma_R^{\rm NL}\sigma_R^2}\right),
\end{aligned}
\end{equation}
where we omitted the $z$ dependence of $\gamma_R,\gamma_R^{\rm NL},\delta_R,\sigma_R$ for the sake of brevity. If we limit ourselves to the first order of the lognormal approximation, $e^{\gamma_R\delta_R}$, this normalization can also be set through $\langle\exp(\gamma_R\delta_R+\mathcal{N}')\rangle=1$, where $\mathcal{N}'(z)=-\gamma_R\sigma_R^2/2$~\cite{Xavier_2016}. This was indeed the choice in Ref.~\cite{Munoz:2023kkg}, where the lognormal approximation was written in terms of $\delta_R'=\delta_R-\gamma_R\sigma_R^2/2$.

\begin{figure}[t!]
    \centering
    \includegraphics[width=\linewidth]{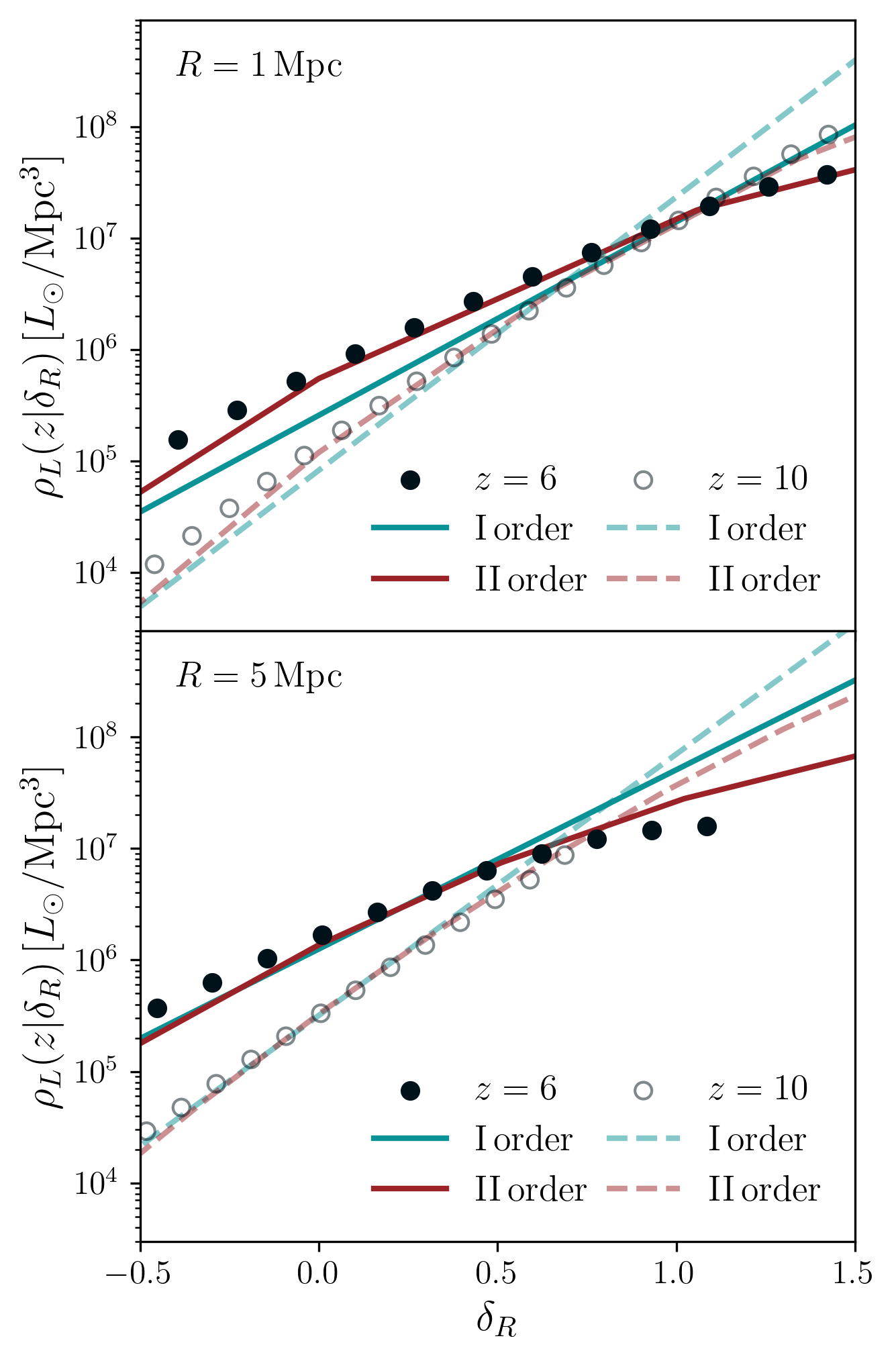}
    \caption{Density-modulated luminosity density $\rho_L(z|\delta_R)$ for different choices of redshift and smoothing radius $R$. The top panel shows $R=1\,{\rm Mpc}$, the bottom $R=5\,{\rm Mpc}$; in both of them, solid lines and filled points are associated with $z=6$, while dashed lines and empty points with $z=10$. While the lines are obtained relying on the first (blue) and second (red) order lognormal approximation in Eq.~\eqref{eq:lognormal}, the points indicate the binned $\rho_L(\vec{x},z)$ in coeval boxes having side $L_{\rm box}=150\,$Mpc and $N_{\rm cell}=$150 cells per side. The boxes are obtained using the cell-by-cell algorithm introduced at the end of Sec.~\ref{sec:line_model}. In all scenarios, the second-order lognormal approximation shows better agreement with the mock coeval boxes, particularly for small $R$, low $z$ or high $\delta_R$.}
    \label{fig:lognormal_approx}
\end{figure}

In Eq.~\eqref{eq:lognormal}, the amplitude of the fluctuations is finally set by the average $\bar{\rho}_L(z)$ in Eulerian space. As anticipated in Eq.~\eqref{eq:phi_LtoE}, its value is estimated by correcting the Lagrangian average $\langle\rho^{\rm Lag}(z)\rangle$ through the factor $\phi_{\rm LroE}(z)$. To estimate it in the context of the lognormal approximation, we require
\begin{equation}
   \phi_{\rm LtoE}(z) = \frac{\langle(1+\delta_R)\exp[\gamma_{R,\rm Lag}\delta_R+\gamma_{R,\rm Lag}^{\rm NL}\delta_R^2]\rangle}{\langle\exp[\gamma_{R,\rm Lag}\delta_R+\gamma_{R,\rm Lag}^{\rm NL}\delta_R^2]\rangle}, 
\end{equation}
where now $\gamma_{R,\rm Lag}$ and $\gamma^{\rm NL}_{R,\rm Lag}$ are the coefficients in Lagrangian space, computed using $\rho^{\rm Lag}(z|\delta_R)$ in Eq.~\eqref{eq:gammas}.\footnote{The Lagrangian coefficients can also be estimated as $\gamma_{R,\rm Lag}\simeq \gamma_R-1$ and $\gamma^{\rm NL}_{R,\rm Lag}\simeq\gamma_R^{\rm NL}+1/2$. While this works well at first order, using the second order requires us to be a bit more precise.} Solving the previous equation returns
\begin{equation}
    \phi_{\rm LtoE}(z)=\frac{1+(\gamma_{R,\rm Lag}-2\gamma_{R,\rm Lag}^{\rm NL})\sigma_R^2}{1-2\gamma_{R,{\rm Lag}}^{\rm NL}\sigma_R^2}, 
\end{equation}
while using the first order approximation, $\phi^{\rm I}_{\rm LtoE}(z)=1+\gamma_{R,\rm Lag}\sigma_R^2$, which is the correction adopted in Ref.~\cite{Munoz:2023kkg}. 

With Eq.~\eqref{eq:lognormal} at hand, we are now in a position to compare the lognormal approximation with calculations in the cell-by-cell coeval boxes described at the end of the previous section. We do so in Fig.~\ref{fig:lognormal_approx} using the luminosity density $\rho_L(z|\delta_R)$; the analogous plot for the SFRD can be produced using both \texttt{oLIMpus} and \texttt{Zeus21}.
The scatter points show the binned $\rho_L(\vec{x},z)$ measured from $z=\{6,10\}$ coeval boxes 
smoothed over $R\,=\,\{1,5\}\,{\rm Mpc}$ in the top and bottom panels respectively.

\noindent The lines show $\rho_L(z|\delta_R)$ computed through the lognormal approximation, up to either the first or second order. From the plot, we can verify that the difference between the first and second order in the lognormal approximation increases for larger $\delta_R$ and smaller $z$, as we expected. The first order approximation generally overestimates the value of $\rho_L(z|\delta_R)$ when $\delta_R$ is large.  
On the other hand, the second order approximation shows very good agreement with the data points under all  conditions.
Therefore, from now on we always adopt the second-order lognormal approximation. In \texttt{oLIMpus}, this is used as the default choice to compute both $\rho_L$ and $\dot{\rho}_{*}$.

{As a final point of this section, we discuss the difference between the expression we obtained for the luminosity density and the one commonly derived using the halo-model approach, which is the standard choice in LIM-related studies~\cite{Bernal:2022jap}. A more detailed discussion can be found in Sect.~\ref{sec:compare_lim}, where we compare the output of \texttt{oLIMpus} with the power spectrum estimated by a halo-model–based code.
The first and most evident difference is that our approach explicitly links the line luminosity to the local density field, whereas the halo model relies on the average halo mass function, similarly to Eq.~\eqref{eq:rho_Lag}. Our choice is crucial for producing maps that consistently trace the underlying density field, the star formation rate, and the line luminosity.
The second difference is the lognormal approximation introduced in this section. Through the definition of the $\gamma$ coefficients in Eq.~\eqref{eq:gammas}, this approach implicitly accounts for the bias (see Sect.~\ref{sec:compare_lim}) between the line luminosity and the underlying density field. In contrast, halo-model approaches generally introduce the line bias by scaling the halo bias with the line luminosity. However, the halo bias is usually calibrated on simulations with a fixed cosmology. While our method still involves an approximation, it is more general. Moreover, because we use a second-order lognormal approximation, the $\gamma$ coefficients capture beyond-linear information on clustering.
Last but not least, the lognormal approach allows us to model both star-forming lines and the IGM 21-cm line consistently, enabling cross-correlation studies.}


\subsection{Line luminosity}\label{sec:line_lum}

Before proceeding, we finally need to characterize the $f(M_h,z)$ function in Eq.~\eqref{eq:rho_Lag}. Since the main focus of \texttt{oLIMpus} are star-forming lines, here we introduce the luminosity-to-halo mass relation, $L(M_h,z)$, required to compute the luminosity density $\rho_L$. In Appendix~\ref{app:SFR} we discuss the star formation rate model that enters the computation of the SFRD, which can be easily modified.

In \texttt{oLIMpus}, we have introduced by default models for OII, OIII, H$\alpha$, H$\beta$, CII, CO line luminosities; these are based on scaling relations calibrated on state-of-the-art simulations, and are summarized in Appendix~\ref{app:lines}. To make the code easily comparable with previous results, we chose to implement models that have been already used in the literature. These, however, were developed by different groups, which implies introducing different parameterizations. We apologize to the reader if this may seem a bit confusing, but at the same time we stress that it is very easy to customize \texttt{oLIMpus} to introduce any other type of model, or even new lines. 

Each model either directly provides $L(M_h,z)$, or introduces a luminosity-to-star formation rate relation, $L(\dot{M}_*,z)$. In the latter case, since for $\dot{M}_*(M_h)$ we use a deterministic function of the host dark matter halo mass (see Appendix~\ref{app:SFR}), the equations directly translate into non-linear $L(M_h)$ relations. 

To account for some level of stochasticity, we introduce the probability distribution function (PDF) $p(L|M_h)$, such that 
\begin{equation}\label{eq:stoch}
\begin{aligned}
    &\tilde{L}(M_h,z) =\int dL'L'p(L'|M_h,z)\\
     &=\int \frac{dL'}{\sqrt{2\pi}\sigma_L} \exp\left[-\frac{(\log L'-\mu(M_h,z))^2}{2\sigma_L^2}\right].
\end{aligned}
\end{equation}
In the second line we used a lognormal PDF having mean $\mu(M_h,z)=\log L(M_h,z)$, i.e.,~we set the deterministic $L(M_h,z)$ to be the median of the distribution. The variance $\sigma^2_L$ is a free parameter; when $\sigma^2_L=0$, the PDF collapses into a Dirac delta centered on $L(M_h,z)$ and we recover the deterministic case. Adding a stochastic contribution shifts the average luminosity density to larger values, because it increases the probability of less-massive halos populating brighter galaxies (see discussion, e.g.,~in Refs.~\cite{Munoz:2023cup, Sun2023_burstiness,Nikolic:2023wea} and Appendix~\ref{app:lines}). 
Moreover, changing the shape of $p(L|M_h)$ e.g.,~based on simulations, potentially offers the opportunity to explore more complex scenarios.  
 
For conciseness in the notation, in the rest of the paper we will always refer to the deterministic $L(M_h,z)$ case, but all derivations are consistent when stochasticity is introduced, and they can be extended straightforwardly. 

Finally, we compute the LIM observable, namely the specific intensity\footnote{The intensity $I_\nu\,[{\rm Jy/sr}]$ is usually used in optical and UV LIM observations. Surveys that access other wavebands, e.g.,~CII in the infrared, more often rely on the brightness temperature
\begin{equation}
    T_\nu\,[\mu K] = \frac{c^3(1+z)^2}{4\pi k_B\nu_{\rm rest}^3H(z)},
\end{equation}
where $k_B$ is the Boltzmann constant.
\texttt{oLIMpus} allows the user to estimate both the observable quantities.}
\begin{equation}\label{eq:Inu}
    I_\nu(z) \,[{\rm Jy/sr}]= \frac{c}{4\pi\nu_{\rm rest}H(z)}\rho_L(z) = c_1^{\rm LIM}(z)\rho_L(z),
\end{equation}
where $c$ is the speed of light and $\nu_{\rm rest}$ the rest-frame frequency of the line.
The previous relation holds both in Lagrangian and Eulerian space, both for the average and for the density-weighted cases. Therefore, the lognormal approximation that we introduced for $\rho_L$ in Eq.~\eqref{eq:lognormal} can be straightforwardly extended to
\begin{equation}\label{eq:Inu_approx}
    I_\nu(\delta_R|z)\simeq c_1^{\rm LIM}(z)c_2^{\rm LIM}(z,R)e^{\gamma_R\delta_R+\gamma_R^{\rm NL}\delta_R^2},
\end{equation}
where in $c_2^{\rm LIM}(z,R)$ we collected all the normalization coefficients, some of which depend on the smoothing scale $R$. This expression will be used in Sect.~\ref{sec:2pt} to construct the  LIM two-point function and power spectrum.


\section{Analytical model for the two-point function}\label{sec:2pt}

The $\rho_L(\delta_R|z)$ and $I_\nu(\delta_R|z)$ lognormal approximations introduced in Eqs.~\eqref{eq:lognormal},~\eqref{eq:Inu_approx}, are the key to develop a fully analytical treatment for the LIM and 21-cm auto- and cross-power spectra. 
To do so in general terms, first of all we consider two lognormal quantities, $a=e^{\gamma_1\delta_{R_1}^2+\gamma_1^{\rm NL}\delta_{R_1}^2}/\mathcal{N}_1$ and $b=e^{\gamma_2\delta_{R_2}^2+\gamma_2^{\rm NL}\delta_{R_2}^2}/\mathcal{N}_2$, each smoothed on a different scale $R_{1,2}$ and normalized to 1. 
We can explicitly derive the two-point function  of these quantities in real space, through\footnote{In Eq.~\eqref{eq:2ptf} the $-1$ term is due to the auto-correlation terms $\langle a^2\rangle$.}
\begin{equation}\label{eq:2ptf}
\begin{aligned}
    \langle ab\rangle &=\frac{\langle\exp(\gamma_1{\delta}_{R_1}+\gamma_1^{\rm NL}{\delta}_{R_1}^2)\exp(\gamma_2{\delta}_{R_2}+\gamma_2^{\rm NL}{\delta}_{R_2}^2)\rangle}{\mathcal{N}_1\mathcal{N}_2}-1\\
    &=-1 + \int \frac{d\delta_{R_1}d\delta_{R_2}}{\sqrt{(2\pi)^2(1-{\rm det}\mathcal{C})}}\exp\left[-\frac{1}{2}\delta^T\mathcal{C}^{-1}\delta\right] \times\\
    &\qquad\times \frac{\langle\exp(\gamma_1{\delta}_{R_1}+\gamma_1^{\rm NL}{\delta}_{R_1}^2)\exp(\gamma_2{\delta}_{R_2}+\gamma_2^{\rm NL}{\delta}_{R_2}^2)\rangle}{\mathcal{N}_1\mathcal{N}_2}
\end{aligned}
\end{equation}
Here, we introduced $\delta^T=\begin{pmatrix} \delta_{R_1} & \delta_{R_2}\end{pmatrix}$ and we defined the covariance matrix 
\begin{equation}
    \mathcal{C}=\begin{pmatrix}
        \sigma_{R_1}^2 & \xi^{R_1R_2}\\
        \xi^{R_1R_2} & \sigma^2_{R_2}
    \end{pmatrix},
\end{equation}
where $\xi^{R_1R_2}=\langle\delta_{R_1}(\vec{x}_1)\delta_{R_2}(\vec{x}_2)\rangle$, which in our case is 
\begin{equation}
\xi^{R_1R_2}(r)=D^2(z)\xi^{R_1R_2}_m(r,z=0),    
\end{equation}
with $r=\vec{x}_2-\vec{x}_1$, $D(z)$ the growth factor and 
\begin{equation}
    \xi_m^{R_1R_2}(r,z=0)={\rm FT}[W(k,R_{1})W(k,R_{2})P_m(k,z=0)]    
\end{equation}
is the matter two-point function at $z=0$ smoothed over scales $R_1$, $R_2$ ($P_m$ is the linear matter power spectrum). Throughout the text, we choose $W(k,R)$ to be a top hat in real space, but its shape can be easily generalized. 

Working through Eq.~\eqref{eq:2ptf}, we can re-express it as
\begin{equation}\label{eq:full_2ptf}
    \langle ab\rangle =\exp\left[\dfrac{N^{R_1R_2}}{D^{R_1R_2}}-\log C^{R_1R_2}\right]{-1}, 
\end{equation}
where 
\begin{equation}
\begin{aligned}
     N^{R_1R_2} &= \gamma_1\gamma_2\xi^{R_1R_2}+\\
     &\quad+\gamma_1^2\sigma_{R_1}^2\left[\frac{1}{2}-\gamma_2^{\rm NL}\sigma_{R_2}^2\left(1-\frac{(\xi^{R_1R_2})^2}{(\sigma_{R_1}\sigma_{R_2})^2}\right)\right]+\\
     &\quad+\gamma_2^2\sigma_{R_2}^2\left[\frac{1}{2}-\gamma_1^{\rm NL}\sigma_{R_1}^2\left(1-\frac{(\xi^{R_1R_2})^2}{(\sigma_{R_1}\sigma_{R_2})^2}\right)\right],\\
     D^{R_1R_2} &=1-2\gamma_2^{\rm NL}\sigma_{R_2}^2-2\gamma_1^{\rm NL}\sigma_{R_1}^2+\\
     &\quad+4\gamma_1^{\rm NL}\sigma_{R_1}^2\gamma_2^{\rm NL}\sigma_{R_2}^2\left(1-\frac{(\xi^{R_1R_2})^2}{\sigma_{R_1}^2\sigma_{R_2}^2}\right),\\
    C^{R_1R_2}&=\sqrt{D^{R_1R_2}}\mathcal{N}_1\mathcal{N}_2.
\end{aligned}
\end{equation}

\begin{figure}[t!]
    \centering
    \includegraphics[width=\linewidth]{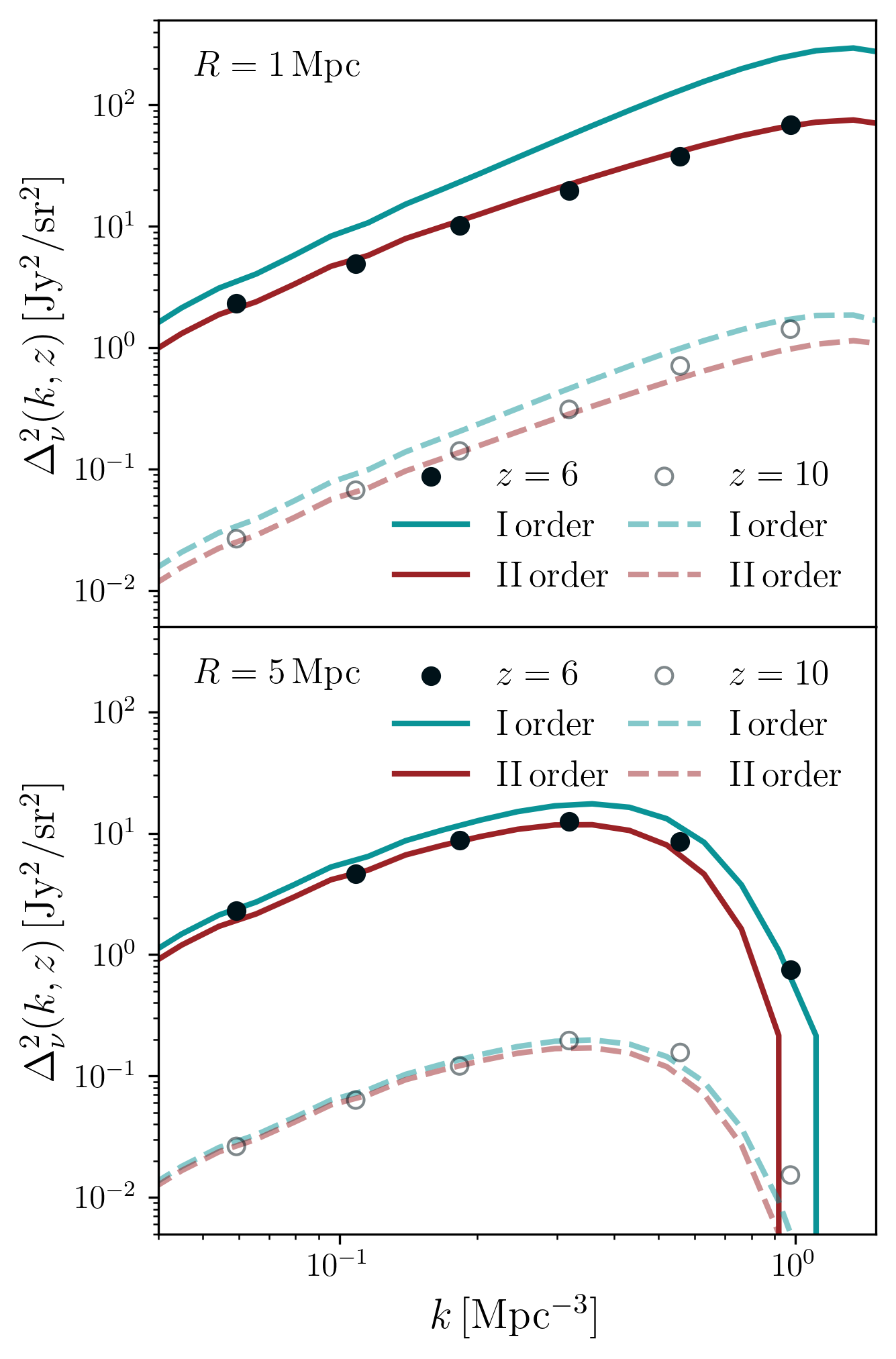}\vspace{-.3cm}
    \caption{Line intensity power spectrum $\Delta_{\nu}^2(k,z)$ for different choices of redshift and smoothing radius $R$. The top panel shows $R=1\,{\rm Mpc}$, the bottom $R=5\,{\rm Mpc}$; the color legend is the same as Fig.~\ref{fig:lognormal_approx}. 
    The points show the power spectrum measured using \texttt{powerbox} on $\{L_{\rm box},N_{\rm cell}\}=\{150\,{\rm Mpc},150\}$ boxes produced through the cell-by-cell algorithm introduced at the end of Sec.~\ref{sec:line_model}. In all scenarios, the second-order lognormal approximation shows remarkable agreement with the mock coeval boxes; the first order case, instead, overestimates the power spectrum on small scales.}
    \label{fig:pk_OIII_scale}
\end{figure}

These lengthy expressions give us a direct way to compute the non-linear power spectrum, by simply specifying the coefficients in Eq.~\eqref{eq:full_2ptf}, scaling the result by the proper amplitude and taking its Fourier transform. In the case of LIM, since line emission is a local process, the auto-power spectrum requires $R_1=R_2=R_0$, so that 
\begin{equation}\label{eq:LIM_Pk}
\begin{aligned}
    P_\nu(k,z) &= {\rm FT}\biggl[\left(c_1^{\rm LIM}(z)c_2^{\rm LIM}(z,R_0)\right)^2\times \\
    &\times \left(\exp(N^{R_0}/D^{R_0} -\log C^{R_0})-1\right)\biggr].
\end{aligned}
\end{equation}
The value of $R_0$ is set either by the minimum scale up to which the approximation can be trusted, or by the resolution of the LIM detector. In the former case, looking at Fig.~\ref{fig:lognormal_approx}, we set $R=1\,{\rm Mpc}$, whereas in the latter the choice depends on the experimental setup. For example, single-dish experiments such as COMAP-EoR~\cite{COMAP:2021nrp} will have a resolution of $\sigma_\theta\sim 4\,{\rm arcmin}/\sqrt{8\log(2)}$, which translates into a comoving scale of $\sim 5$\,Mpc at $z\sim 7$. 

In Fig.~\ref{fig:pk_OIII_scale} we show an example power spectrum, defined as $\Delta_\nu^2(k,z)=k^3P_\nu(k,z)/(2\pi^2)$ for different redshifts and smoothing radii. Similarly to Fig.~\ref{fig:lognormal_approx}, we compare our baseline analytical power spectrum with the power spectrum estimated using \texttt{powerbox}\footnote{\url{https://github.com/steven-murray/powerbox/tree/main}}~\cite{Murray2018} on the same cell-by-cell calculated boxes used in Fig.~\ref{fig:lognormal_approx}. The second order lognormal approximation shows remarkable agreement with the cell-by-cell mock coeval boxes, the largest difference being found on scales close to the smoothing radius, when this is large. The amplitude offset between the first order and second order power spectra on large scales (low $k$) is due to the Lagrangian-to-Eulerian correction factor $\phi_{\rm LtoE}$ in Eq.~\eqref{eq:phi_LtoE}: since $\gamma_R^{\rm NL} \neq \gamma_{R,\rm Lag}^{\rm NL}$, setting $\gamma_R^{\rm NL}=0$ as in the limit of large scales does not return the same power spectrum as in the first order computation. 

Equation~\eqref{eq:LIM_Pk} can be straightforwardly extended to the computation of the cross-power spectrum,\footnote{{Eq.~\eqref{eq:LIM_cross} computes the cross power spectrum between star forming lines. To compute their cross-spectrum with 21-cm, the equation has to be extended; in fact, the correlation of the 21-cm signal depends on the underlying auto- and cross- correlations between the density field and the star-formation rate dependent quantities, namely the Lyman-$\alpha$ and X-ray fluxes (as detailed in Refs.~\cite{Munoz:2023kkg,Cruz:2024fsv}) and the ionized fraction during the EoR~\cite{Sklansky:2025}.}}
\begin{equation}\label{eq:LIM_cross}
\begin{aligned}
    &P_{\nu_1\nu_2}(k,z) = {\rm FT}\biggl[c_{1,\nu_1}^{\rm LIM}(z)c_{2,\nu_1}^{\rm LIM}(z,R_1)c_{1,\nu_2}^{\rm LIM}(z)c_{2,\nu_2}^{\rm LIM}(z,R_2)\times \\
    &\qquad\times \left(\exp(N^{R_1R_2}/D^{R_1R_2} -\log C^{R_1R_2})-1\right)\biggr].
\end{aligned}
\end{equation}

The previous equation can easily account for two different smoothing radii; this will be useful when cross-correlating measurements from different instruments. 

\begin{figure*}[ht!]
    \centering
    \includegraphics[width=\linewidth]{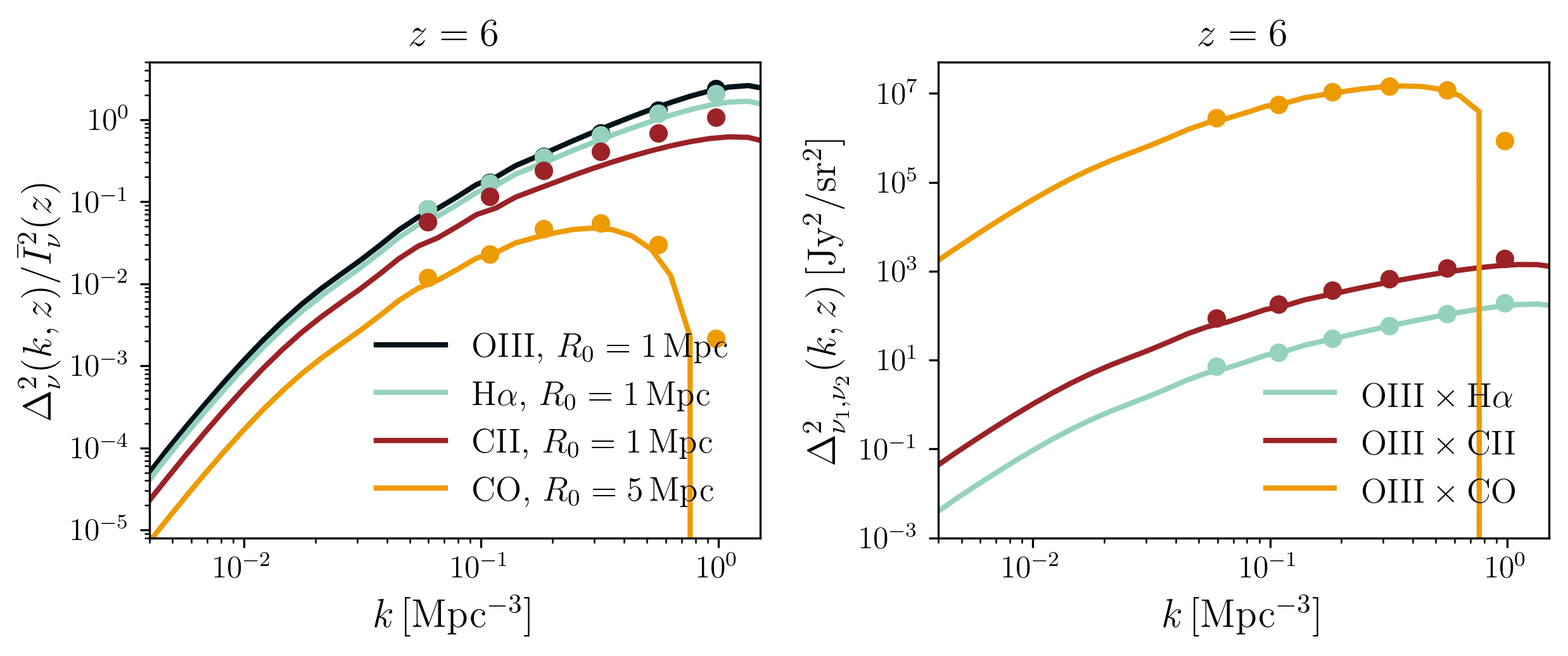}
    \caption{Left: normalized {auto}-power spectra for OIII, H$\alpha$, CII and CO(2-1), each of which is divided by the observed intensity $\bar{I}_\nu^2(z)$. Except for OIII and H$\alpha$, \texttt{oLIMpus} uses different models for different lines (see Appendix~\ref{app:lines}). We made this choice in order to include by default in our code models that are actively used in the literature, each of which is calibrated on state-of-the art simulations; other models can be easily customized into the code.
    The choice of a larger smoothing radius $R_0$ for CO is illustrative. Right: cross power spectra between OIII and the other lines at $z=6$. In both plots, the points are obtained using \texttt{powerbox} on \{$L_{\rm box}, N_{\rm cell}$\}=\{150\,Mpc, 150\} boxes produced with the cell-by-cell alghorithm described at the end of Sect.~\ref{sec:line_model}.}
    \label{fig:cross}
\end{figure*}
We show in Fig.~\ref{fig:cross} some examples of the auto- and cross- power spectra that can be produced using \texttt{oLIMpus}; in the auto- case, we normalized each power spectrum by dividing by $\bar{I}_\nu^2(z)$. For illustration, we choose different values of $R_0$ for different lines. 

Finally, we need to account for distortions in the power spectrum caused by the peculiar velocities of luminous sources, which perturb the observed positions of the emitters. These lead to the so-called redshift-space distortions (RSD),\footnote{For the time being, we neglect distortions arising from general relativistic effects.} which we introduce in our formalism following the standard derivation~\cite{Kaiser_1987,peebles_1980,Dodelson:2003ft}, summarized in Appendix~\ref{app:RSD}. In short, the LIM auto-power spectrum in redshift space becomes
\begin{equation}\label{eq:pLIM_rsd}
\begin{aligned}
    &P^{\rm RSD}_\nu(k,z)=P_\nu(k,z)+\bar{I}_\nu^2(z)f^2(z)\mu^4P_m(k,z)\\
    &\qquad\qquad\qquad +2f(z)\mu^2\bar{I}_\nu(z) P_{\nu m}(k,z),
\end{aligned}
\end{equation}
where all the new terms have $\bar{I}_\nu^2$ dimension ($P_{\nu m}$ already contains a power of $\bar{I}_\nu$). In the previous expression, $\mu$ is the cosine of the angle with respect to the line-of-sight. In the case of cross-correlation between two lines, the power spectrum turns out to be
\begin{equation}\label{eq:pLIM_rsd_cross}
 \begin{aligned}
   &P^{\rm RSD}_{\nu_1\nu_2}(k,z)=P_{\nu_1\nu_2}(k,z)+\bar{I}_{\nu_1}(z)\bar{I}_{\nu_2}(z)f^2(z)\mu^4P_m(k,z)\\
    &\qquad +f(z)\mu^2\bar{I}_{\nu_1}(z) P_{\nu_1 m}(k,z)+f(z)\mu^2\bar{I}_{\nu_2}(z) P_{\nu_2 m}(k,z).   
\end{aligned}
\end{equation}
As discussed in Appendix~\ref{app:RSD}, by further scaling $P^{\rm RSD}_{\nu,\nu_1\nu_2}(k,z)$ by $1/[1+(k\mu\sigma_{\rm FoG})^2/2]^2$ we can also account for the Fingers-of-God effect~\cite{Peacock:1993xg,Ballinger:1996cd}, which damps the power spectrum on small scales along the line-of-sight.

To conclude, we briefly mention that our formalism can be extended to the computation of the 21-cm power spectrum; the result, however, is much more complicated, since it requires to combine different non-local and non-linear functions of the SFRD. We refer the interested reader to the original \texttt{Zeus21} papers, Refs.~\cite{Munoz:2023kkg,Cruz:2024fsv}. We consistently included the 21-cm signal computation in \texttt{oLIMpus} through the \texttt{Zeus21} submodule, where we improved the implementation by including the second order in the logarithmic approximation of the SFRD, similarly to Eq.~\eqref{eq:lognormal}. As already mentioned, in our upcoming work~\cite{Libanore:2025gte}, we will explore the cross-correlation between 21-cm and the star-forming lines.



\subsection{Shot Noise}\label{sec:shot_noise}

The LIM signal is produced by a discrete population of sources that resides inside dark matter halos; because of this, on top of the clustering, the LIM power spectrum contains a scale-independent contribution, usually indicated as shot noise or Poisson noise. 

To introduce the shot noise in our formalism, we follow the standard derivation (e.g.,~Ref.~\cite{peebles_1980}) and we extend it to the luminosity density $\rho_L$; the same procedure holds for the SFRD. First of all, we define $\bar{n}=\int d\log M_h \,dn/d\log M_h$ as the halo average number density in units of inverse volume, and the local number density as
\begin{equation}
    n(\vec{x})=\bar{n}(1+\delta(\vec{x}))=\sum_i\delta^D(\vec{x}-\vec{x}_i),
\end{equation}
where in the second equality we are summing over halo positions $\vec{x}_i$. We invert this equation to get $\delta(\vec{x})$ and its discrete Fourier transform 
\begin{equation}\label{eq:delta_k_sn}
\begin{aligned}
    \tilde{\delta}(\vec{k}) &= \frac{1}{V}\int d^3x\left(\frac{1}{\bar{n}}\sum_i\delta^D(\vec{x}-\vec{x}_i)-1\right)e^{-i\vec{k}\cdot\vec{x}}\\
    &= \frac{1}{\bar{n}V}\sum_j\mathcal{N}_je^{-i\vec{k}\cdot\vec{x}_j}-\delta^D(\vec{k}),
\end{aligned}
\end{equation}
where we switched the sum over halo positions into a sum over infinitesimal cells, each containing either 1 or 0 halos; their number is expressed by $\mathcal{N}_j=\{0,1\}$. 
This expression relies on the number density of the full halo distribution, without accounting for mass dependence. 

To introduce it, we separate $n(\vec{x})$ into the sum of infinitesimal mass-bin contributions, $n(\vec{x})=\sum_m n(\vec{x}|M_m)$. For each $m$ bin separately, it holds $n_m(\vec{x}) = \bar{n}_m(1+\delta(\vec{x}))$, $\bar{n}_m$ being the number of halos expected from the HMF in the mass bin per unit volume. In analogy to the previous case, in each bin we can apply the Fourier transform and define $\tilde{\delta}_m(\vec{k})$; summing their contributions, we introduce 
\begin{equation}
\begin{aligned}
    \tilde{\rho}_h(\vec{k})&= \sum_m \tilde{\delta}_m(\vec{k}) = \sum_m \frac{1}{\bar{n}_mV}\sum_j \mathcal{N}_{j,m}e^{-i\vec{k}\cdot\vec{x}_j},\\
    &=\int d\log M_h\frac{1}{{dn}/{d\log M_h}\, V}\sum_j\mathcal{N}_{j} e^{-i\vec{k}\cdot\vec{x}_j},
\end{aligned}
\end{equation}
The luminosity density can be written in analogy to this quantity; in particular, we need to redefine the infinitesimal cells $\mathcal{N}^{L}_{j,m}$, so that they contain either 1 or 0 sources, each contributing to the average luminosity as $\langle \mathcal{N}_m^{L}\rangle=\bar{n}_m\Delta V\,\bar{\rho}_L(M_m)$, where $\bar{\rho}_L(M_m)$ is the average Eulerian luminosity density in the bin. In this case, $\langle (\mathcal{N}_m^{L})^2\rangle = \bar{n}_m\Delta V \langle{\rho^2_L}(M_m)\rangle$, where only in this case we express the Eulerian mean through $\langle \rangle$. Thanks to this definition, we can write 
\begin{equation}
\begin{aligned}
    \tilde{\rho}_{L}(\vec{k})&=\int d\log M_h\frac{1}{{dn}/{d\log M_h}\, V}\sum_j\mathcal{N}_{j}^{L}(M_h) e^{-i\vec{k}\cdot\vec{x}_j},
\end{aligned}
\end{equation}
and use this expression to estimate the shot noise power spectrum\footnote{In the standard derivation of the halo shot noise power spectrum, $\tilde{\delta}(\vec{k})$ in Eq.~\eqref{eq:delta_k_sn} is used to get
\begin{equation}
    \begin{aligned}
        P_\delta(k,z)&=V\langle|\delta(\vec{k},z)^2|\rangle\\
        &=V\langle\frac{1}{\bar{n}V}\sum_j\mathcal{N}_je^{-i\vec{k}\cdot\vec{x}_j}\frac{1}{\bar{n}V}\sum_i\mathcal{N}_ie^{i\vec{k}\cdot\vec{x}_i}\rangle\\
        &=\frac{1}{\bar{n}^2V}\bar{n}V + \frac{1}{\bar{n}^2V}\sum_{i\neq j}\bar{n}^2V^2\xi_{ij}e^{-i\vec{k}(\vec{x}_j-\vec{x}_i)}\\
        &= P_{\rm shot}(z)+ P_h(k,z)
    \end{aligned}
\end{equation}
where we neglected the $\vec{k}=0$ monopole term, we used $\langle\mathcal{N}^2\rangle=\langle\mathcal{N}\rangle=\bar{n}V$ and in the last line we defined $\vec{r}=\vec{x}_j-\vec{x}_i$ and we changed the sum over $i\neq j$ positions into an integral over $\vec{r}$ to get the halo two-point function $\xi_{ij}$. The derivation in the main text follows the same logic, omitting the clustering term for brevity.} as
\begin{equation}
    \begin{aligned}
        P_{\rho_{L}}&(k,z)=V\langle|\tilde{\rho}_L(\vec{k})|^2\rangle\\
        &=V\int d\log M_h\frac{1}{\bar{n}^2V^2}\sum_{i=j}\langle \mathcal{N}_j^{L}\mathcal{N}_i^{L}e^{-i\vec{k}(\vec{x}_j-\vec{x}_i)}\rangle\\
        &=\int d\log M_h\frac{1}{\bar{n}^2V}\langle{\rho}_L^2(M_h)\rangle\bar{n}V\\
        &= \int d\log M_h\frac{dn}{d\log M_h}\langle L^2(M_h,z)\rangle,
    \end{aligned}
\end{equation}
where after the third line we expressed $\rho_L(M_h)$ in terms of the luminosity per halo, using the HMF as PDF. Scaling this result to the units of the specific intensity, we get the shot noise contribution to the LIM power spectrum,
\begin{equation}\label{eq:shot}
    P_{\rm shot}(z)=(c_1^{\rm LIM}(z))^2\int d\log M_h\frac{dn}{d\log M_h}\langle L^2(M_h,z)\rangle.
\end{equation}
For the cross-power spectrum, we set $P_{\rm shot}(z)=0$. Finally, we multiply $P_{\rm shot}(z)$ by $W^2(k,R)$ to get a smoothing scale consistent with the clustering power spectrum. {It is worth noting that this shot-noise expression applies only to lines produced inside galaxies. Shot noise also affects the 21-cm power spectrum, adding small-scale power to the cross-correlation of the star-formation–related quantities that enter its definition~\cite{Munoz:2023kkg}. Furthermore, since neutral hydrogen resides both inside galaxies and in the IGM, the corresponding shot noise cross spectrum between these two components is non-zero. We will investigate these aspects in future work. }

\begin{figure}[ht!]
    \centering
    \includegraphics[width=\linewidth]{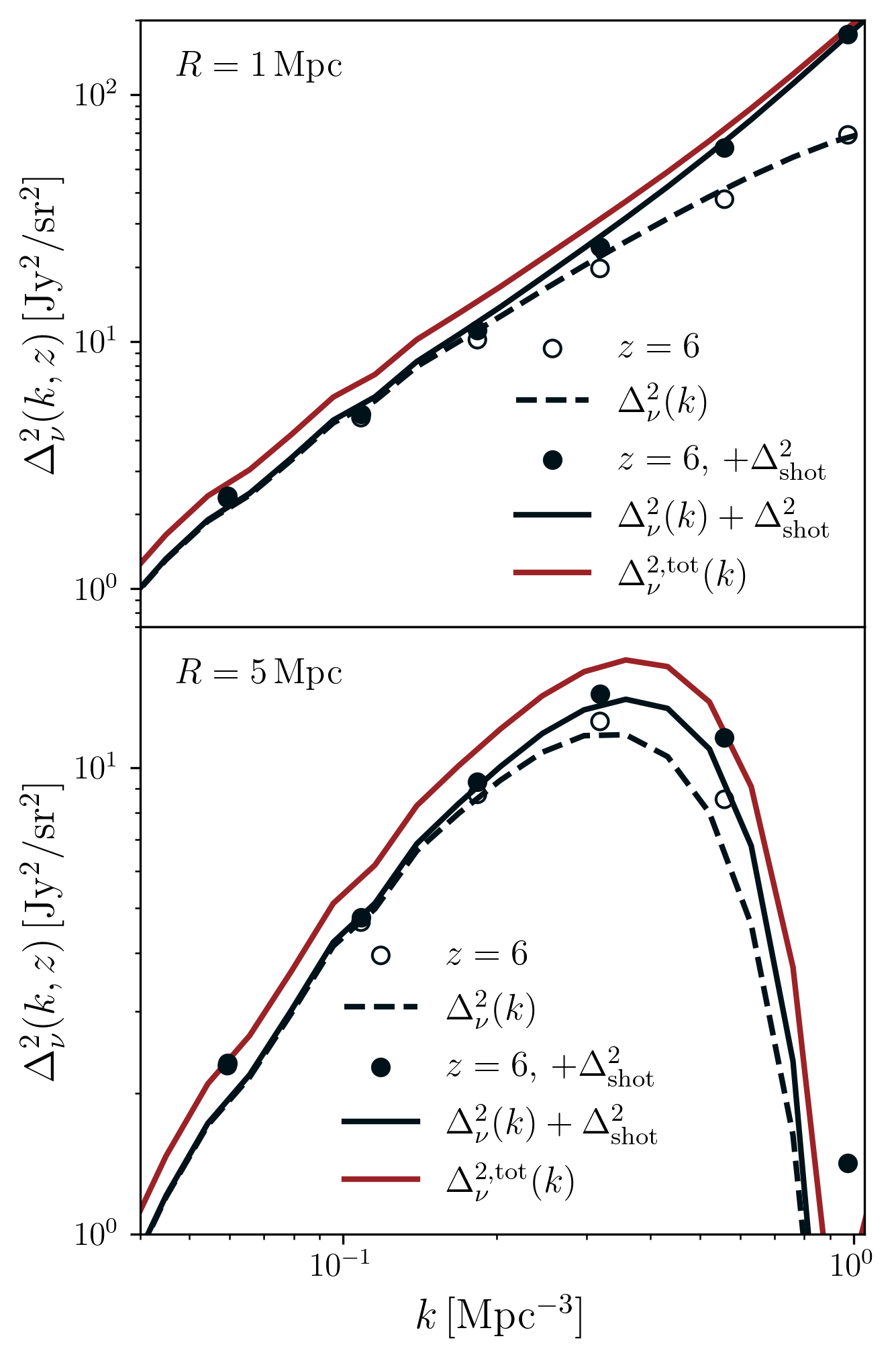}
    \caption{LIM power spectrum with (solid and black dots) and without (dashed and white dots) shot noise contribution, see Sect.~\ref{sec:shot_noise}. The red lines include spherically-averaged RSD, see Appendix~\ref{app:RSD}. The points are measured with \texttt{powerbox} in coeval boxes $L_{\rm box}=150\,$Mpc and $N_{\rm cell}=$150. For the white points, we use the intensity box produced with the cell-by-cell algorithm at the end of Sec.~\ref{sec:line_model}; the black points, instead, are measured from a box produced by summing the intensity box with the shot noise box, realized as a Gaussian field in \texttt{powerbox} based on $P_{\rm shot}(z)$ in Eq.~\eqref{eq:shot}; the smoothing over different radii (top, $R=1\,$Mpc, bottom $R=5\,$Mpc) is done a posteriori. The cell-by-cell algorithm does not include RSD.}
    \label{fig:shot_noise_and_RSD}
\end{figure}

{With respect to star forming lines, therefore, }the complete output of \texttt{oLIMpus} can be written in real space as
\begin{align}
    {\Delta}^2_\nu(k,z)&=\frac{k^3}{2\pi^2}(P_{\nu}(k,z)+W^2(k,z)P_{\rm shot}(z))\,\,\,\,{(\rm auto)} \nonumber\\
    {\Delta}^2_\nu(k,z)&=\frac{k^3}{2\pi^2}P_{\nu_1\nu_2}(k,z)\,\,\,\,{(\rm cross)}
\end{align}
or in redshift space
\begin{align}
    \tilde{\Delta}^2_\nu(k,z)&=\frac{k^3}{2\pi^2}(P^{\rm RSD}_{\nu}(k,z)+W^2(k,z)P_{\rm shot}(z))\,\,\,\, {(\rm auto)} \nonumber \\
    \tilde{\Delta}^2_\nu(k,z)&=\frac{k^3}{2\pi^2}P^{\rm RSD}_{\nu_1\nu_2}(k,z)\,\,\,\,{(\rm cross)}.
\end{align}
The latter can also be scaled by the Fingers-of-God contribution to account for non linearity in the RSD on small scales (see Appendix~\ref{app:RSD}).

Figure~\ref{fig:shot_noise_and_RSD} compares the LIM power spectrum obtained from \texttt{oLIMpus} with and without shot noise. As expected, the scale-independent contribution of the shot noise becomes relevant on small scales; when the smoothing scale is too large, the shot noise gets damped similarly to the clustering term. In the plot, we also show the power spectrum measured from the coeval boxes produced with the cell-by-cell algorithm, to which we summed an uncorrelated Gaussian field produced by passing $P_{\rm shot}(z)$ as input to \texttt{powerbox}. Alternatively, the shot noise could be numerically generated by sampling discrete sources in the box; since this would make the implementation more computationally expensive, for the aim of this work, we decided to rely on the analytical case. Meanwhile, we are testing the second-order lognormal approximation and its implications directly against N-body simulations~\cite{Ventura:2025}. 

Moreover, in the figure we compare the real space $\mu=0$ and redshift space $\mu = 0.6$ power spectra (see details in Appendix~\ref{app:RSD}). Overall, RSD slightly increase the amplitude on all scales; where the shot noise dominates, the effect becomes negligible. 
The red solid line is the complete redshift-space LIM power spectrum $\tilde{\Delta}_\nu^{\rm tot}(k,z)$ expected from \texttt{oLIMpus}. 


\subsection{Fast exploration of the parameter space}\label{sec:science}

Thanks to its short computational time, \texttt{oLIMpus} can be used to explore how LIM observables depend on various parameters which are used to model the star formation rate, the line luminosity or the underlying cosmological framework. The auto- and cross- power spectra of single and multiple lines can be produced consistently, and a large parameter space can be explored efficiently. Because of this, \texttt{oLIMpus} will be well suited to perform Bayesian inference analyses on upcoming LIM data.

\begin{figure*}
    \centering
    \includegraphics[width=.9\linewidth]{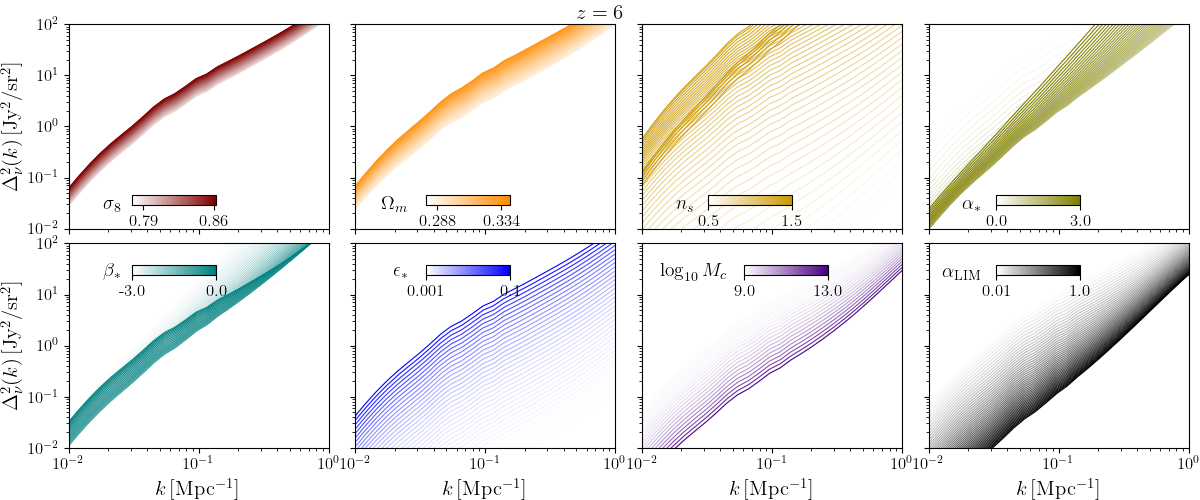}
    \includegraphics[width=.9\linewidth]{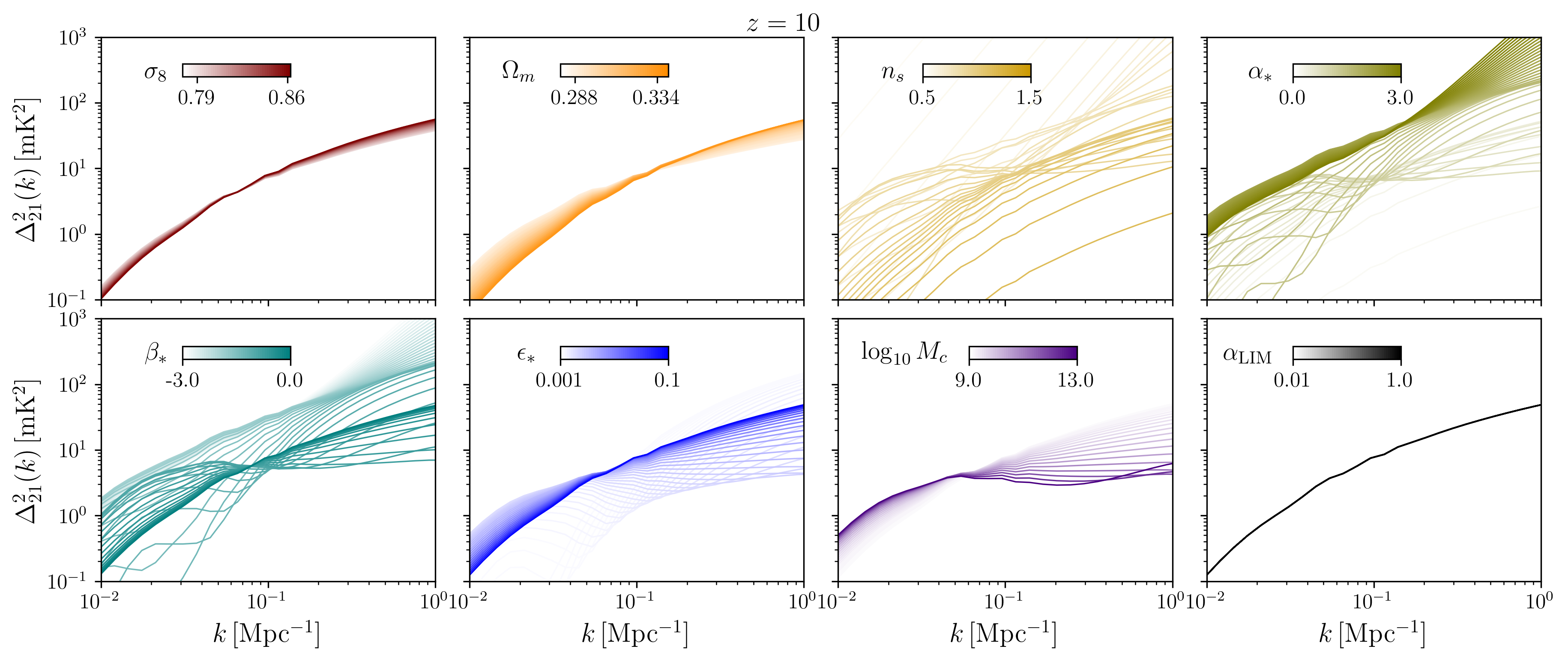}
    \caption{Auto-power spectra of star-forming lines ($z=6$, panels one to eight) and 21-cm ($z=10$, panels nine to sixteen) when a series of astrophysical and cosmological parameters are varied; in each panel we vary only one parameter, while fixing the others to their fiducial values. Left to right and top to bottom, the panels show variations in: (i) cosmology: through the clustering amplitude $\sigma_8$; the matter density $\Omega_m$; the tilt of the primordial power spectrum $n_s$; (ii) The SFR model: through the coefficients suppressing the low halo mass ($\alpha_*>$0) and high halo mass ($\beta_*<$0) tails of the star formation rate power law; the star formation rate efficiency $\epsilon_*$; the turn over mass $M_c$ (see Eq.~\eqref{eq:fstar}; note that we use a different sign convention respect to Ref.~\cite{Sabti:2021xvh}); (iii) The star-forming line properties: through the $\alpha_{\rm LIM}$ coefficient in Eq.~\eqref{eq:L}; this does not affect the 21-cm power spectrum. These plots can be compared with Fig.~2 in Ref.~\cite{Sabti:2021xvh} to understand how LIM surveys will provide complementary information to the galaxy UV luminosity function measured by HST or JWST, see the main text for details.}
    \label{fig:var_par}
\end{figure*}

LIM surveys are sensitive to the full population of sources, including the faintest ones; their measurements are hence complementary to the galaxy ultraviolet (UV) luminosity function, which only captures the brightest galaxies. To illustrate this point, in Fig.~\ref{fig:var_par} we show how the LIM power spectrum changes as a function of a set of cosmological and astrophysical parameters. We show both changes in the power spectrum of a star-forming line (OIII) and of the 21-cm line at higher $z$; the latter is obtained thanks to the \texttt{Zeus21} submodule interfaced with \texttt{oLIMpus}. The plot can be directly compared with Fig.~2 in Ref.~\cite{Sabti:2021xvh}, where the $z=6$ UV luminosity function is shown as a function of the same set of parameters. 

Some interesting aspects can be noted in Fig.~\ref{fig:var_par}. First of all, the parameters $\alpha_*$ and $\beta_*$ control, respectively, the SFR suppression on low and high halo masses (Eq.~\eqref{eq:fstar}); consequently, when varying them, the star-forming line power spectrum is mostly affected at small and large scales, respectively. On the contrary, the situation for 21-cm is more complicated, due to the different capability the sources have in affecting locally and non-locally the evolution of the gas temperature in the IGM. 

Secondly, the parameter $n_s$ has a stronger impact in the LIM signal compared with the UV luminosity function, since it mostly affects the matter power spectrum on small scales, hence varying the abundance of small-size dark matter halos hosting faint sources. 

Lastly, small values of $M_c$ boost the relative importance of small halos and faint sources compared to the more massive and rarer halos that host the brightest galaxies. This boosts the amplitude of the LIM signal, but at the same time it suppresses the UV luminosity function of bright sources. 

Finally, we stress how, from the figure, a high degeneracy emerges among all these parameters, which makes it even more important to include the full parameter set in future data analyses. To do so, a fast and efficient code such as \texttt{oLIMpus} is crucial.

\section{Comparison with other codes}\label{sec:compare_codes}

In this Section, we want to test our approach by comparing \texttt{oLIMpus} outputs with results from other publicly-available codes that can estimate the LIM power spectrum during the EoR. 
In particular,
we use the following public codes:\footnote{The goal of this section is to discuss some of the features of these codes, with the main goal of comparing them with \texttt{oLIMpus}. We do not intend to provide an exhaustive description of their formalism or implementation; instead, we refer the interested reader to the original publications for comprehensive discussions.}
\begin{itemize}[itemsep=1pt,topsep=5pt]
    \item \texttt{lim}\footnote{\url{https://github.com/jl-bernal/lim}}~\cite{Bernal:2019jdo}: a fully-analytical code that computes the LIM power spectrum using scaling relations for $L(M_h,z)$ and relying on the underlying HMF and $P_m(k,z)$, which is computed from \texttt{CLASS}. 
    \item \texttt{LIMFAST}\footnote{\url{https://github.com/lluism/LIMFAST}}~\cite{Mas-Ribas:2022jok,Sun:2022ucx,Sun:2024vhy}: this code builds on \texttt{21cmFAST}, extending it in different aspects. First, it introduces the more detailed galaxy formation model from Ref.~\cite{Furlanetto_2021}; this is used to estimate the radiation fields based on stellar spectral energy distributions (SEDs) tabulated from Ref.~\cite{BPASS_2017} for different metallicities. With this approach, the code coevolves the galaxy population and reionization. 
    \item \texttt{21cmFASTv4}\footnote{\url{https://github.com/21cmfast/21cmFAST}}~\cite{Davies:2025wsa}: this new release of \texttt{21cmFAST}, besides improving the star formation model, introduces a discrete source model, in which halos and galaxies are stochastically sampled from the conditional HMF and from semi-empirical relations. Line luminosities are then calculated in post-processing on the sampled galaxy field.
\end{itemize}

\subsection{Comparison to \texttt{lim}}\label{sec:compare_lim}

First of all, it is interesting and informative to compare our formalism with the analytical approach of \texttt{lim}, {based on the halo model approach. In this code,} the line intensity power spectrum is defined as 
\begin{equation}\label{eq:lim_pk_bernal}
\begin{aligned}
    &P_\nu(k,z)=\langle{I}_\nu(z)\rangle^2\bar{b}_\nu^2(z)D^2(z)P^{\rm lin}_m(k,z=0).\\
\end{aligned}
\end{equation}
In the previous equation, we only included the so-called two-halo clustering term, since this is the one accounted for in our formalism; \texttt{lim}, however, also allows the user to include the one-halo term to model intra-halo correlations that boost the power on small scales~\cite{Smith:2002dz,Mead:2020vgs}. We omit the one-halo contribution for the moment, while planning to explore it in a follow-up work.

In Eq.~\eqref{eq:lim_pk_bernal}, following the notation of Sect.~\ref{sec:line_model} we indicated the mean intensity as $\langle I_\nu(z)\rangle$, since \texttt{lim} computes it by integrating over the average HMF, without correcting between Lagrangian and Eulerian space. This implies a small offset compared to the amplitude of our power spectrum, of the order of $\phi_{\rm LtoE}$ in Eq.~\eqref{eq:phi_LtoE}.

On the other hand, the average linear bias $\bar{b}_\nu$ is defined based on the Eulerian halo bias $b_h(M_h,z)$~\cite{Tinker:2010my}, as 
\begin{equation}\label{eq:bias_bernal}
    \bar{b}_\nu(z)=\frac{\int dM_hL(M_h,z)b_h(M_h,z)dn(z)/dM_h}{\int dM_hL(M_h,z)dn(z)/dM_h}.
\end{equation}
The power spectrum in Eq.~\eqref{eq:lim_pk_bernal} is then scaled with the RSD contribution and summed to the shot noise, to get
\begin{equation}\label{eq:lim_pk_bernal_tot}
\begin{aligned}
    \tilde{P}_\nu(k,z)&= \left[\frac{1+{f(z)}\mu^2/{\bar{b}_{\nu}(z)}}{1+\left({k_\parallel\sigma_{\rm FoG}}/{2}\right)^2}\right]^{2}P_\nu(k,z) + P_{\rm shot}(z),\\
    P_{\rm shot}(z)&=(c_1^{\rm LIM}(z))^2\int dM_h \frac{dn}{dM_h}(z)L^2(M_h,z).
\end{aligned}
\end{equation}
In the RSD term, the numerator represents the Kaiser term, while the denominator introduces the Lorentzian Fingers-of-God effect (FoG), relevant on small scales along the LoS modes (see Appendix~\ref{app:RSD} for detail).

The quantities that enter Eqs.~\eqref{eq:lim_pk_bernal},~\eqref{eq:bias_bernal},~\eqref{eq:lim_pk_bernal_tot} can be remapped in terms of our formalism. In fact, if we limit the lognormal approximation of to the first order, $e^{\gamma_R\delta_R}$, the two-point function reads as $\xi^{R_0R_0}=e^{\gamma_{R_0}^2\xi_m^{R_0R_0}}-1$. By taking the linear-order Taylor expansion and applying the Fourier transform,  we write the LIM power spectrum as $P^{\rm lin}_\nu(k,z)\simeq \bar{I}_\nu^2\gamma_{R_0}^2W^2(k,R_0)P^{\rm lin}_m(k,z)$.
Comparing this with Eq.~\eqref{eq:lim_pk_bernal}, it is evident that $\gamma_R$ plays the role of the linear bias factor. This is reasonable, since we defined $\gamma_R$ as the derivative of the density-weighted luminosity density, hence it encodes information both on the line emission and on the growth of structure. 
However, once we introduce the second order in the lognormal, Taylor expanding Eq.~\eqref{eq:full_2ptf} at linear order  yields
\begin{equation}\label{eq:linear_second_order}
    P^{\rm lin}_\nu(k,z)\simeq \bar{I}_\nu^2\left(\frac{\gamma_{R_0}}{1-2\gamma_{R_0}^{\rm NL}\sigma_{R_0}^2}\right)^2W^2(k,R_0)P_m^{\rm lin}(k,z).
\end{equation}
Therefore, in our case, the ``bias" contains information beyond the linear level; since $\gamma_{R_0}^{\rm NL}<0$, the non-linear correction decreases the amplitude of the power spectrum with respect to the \texttt{lim} formalism. This comparison helps understanding the role of $\gamma$ and $\gamma_{R_0}^{\rm NL}$, but we recall that it is not used in our code, where instead we rely on the full expression in Eq.~\eqref{eq:full_2ptf}.

As for the RSD, the Kaiser term in Eq.~\eqref{eq:lim_pk_bernal_tot} can be rewritten as $P_\nu^{\rm RSD}(k,z)=P_\nu(k,z)+f^2\mu^4\langle{I}_\nu\rangle^2P_m(k,z)+2f\mu^2\langle{I}_\nu\rangle^2\bar{b}_\nu P_m(k,z)$. In the linear approximation where $\bar{b}_\nu\simeq \gamma_{R}$, this expression is analogous to our formalism, Eq.~\eqref{eq:pLIM_rsd} (modulus the small-amplitude correction due to $\phi_{\rm LtoE}$). 
The Lorentzian FoG effect is the same between \texttt{oLIMpus} and \texttt{lim}.

Finally, regarding the shot noise, the \texttt{lim} expression is analogous to ours in Eq.~\eqref{eq:shot}; both describe this contribution in terms of the second moment of the luminosity.

\begin{figure}
    \centering
    \includegraphics[width=\linewidth]{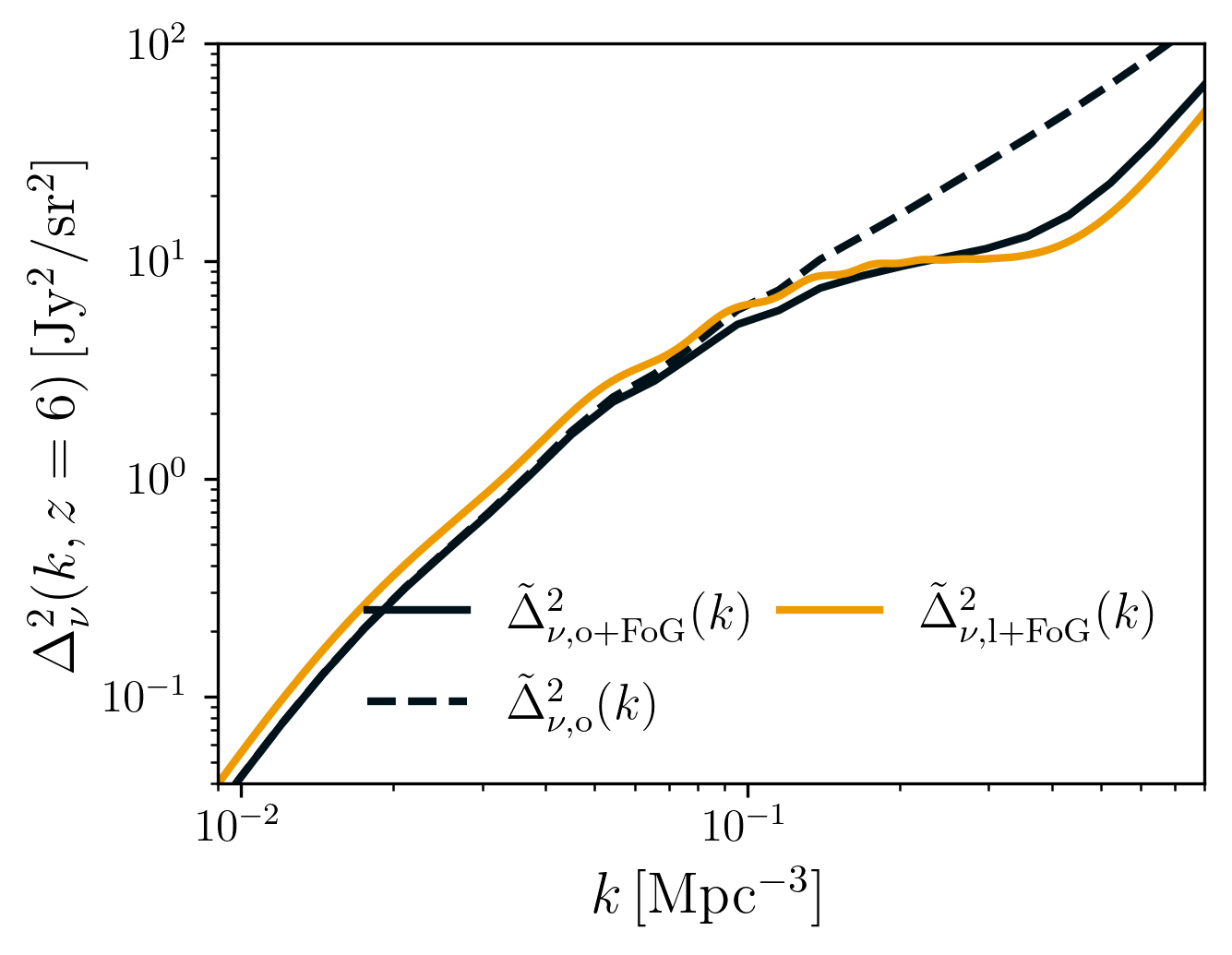}
    \caption{Comparison between the LIM power spectrum from \texttt{lim} ($\tilde{\Delta}^{2}_{\nu,\rm l+FoG}$, yellow) and from \texttt{oLIMpus}, with ($\tilde{\Delta}^{2}_{\nu,\rm o+FoG}$, black) and without ($\tilde{\Delta}^{2}_{\nu,\rm o}$, black dashed) Figers-of-God. All the lines include shot noise and use the spherically-averaged value $\mu = 0.6$. For the FoG contribution, \texttt{lim} and \texttt{oLIMpus} both use $\sigma_{\rm FoG}=7\,{\rm Mpc}$. As described in the main text, the amplitude offset at large scales is due to $\phi_{\rm LtoE}$ and to the $\gamma_R^{\rm NL}$ contribution in Eq.~\eqref{eq:linear_second_order}. 
    The drop of amplitude on small scales, instead, is due to the FoG. 
    } 
    \label{fig:compare_lim}
\end{figure}

Figure~\ref{fig:compare_lim} summarizes the discussion we made so far by comparing the LIM power spectra of \texttt{oLIMpus} and \texttt{lim}; to produce the plot, we assumed the same HMF, $\dot{M}_*(M_h)$, and $L(M_h,z)$ in both codes. 

\subsection{Comparison to \texttt{LIMFAST} and \texttt{21cmFASTv4}}

The comparison between our results and the outputs of \texttt{LIMFAST} is less straightforward. First of all, the code evolves the density and velocity fields using the second-order Lagrangian perturbation theory (2LPT, see e.g.,~Refs.~\cite{Bouchet:1994xp}), which both takes care of non-linear cosmological evolution and of the Lagrangian-to-Eulerian conversion. Moreover, since \texttt{LIMFAST} numerically accounts for cell displacement due to the velocity field~\cite{Jensen_2013}, it introduces RSD in the quasi-linear regime. Most importantly, the line emission in \texttt{LIMFAST} is modeled by using tabulated SEDs. Finally, the astrophysical model includes more dependencies than the ones we relied on in this work, e.g.,~on metallicity and stellar feedback. {In principle, these contributions could be incorporated into our code. One approach is to model an analytical functional form that captures their effect and include it in the same way as the stochastic term in Eq.~\eqref{eq:stoch}; for example, the model in Ref.~\cite{Liu_2024} could be used to introduce the effect of bursty star formation induced by delayed supernova feedback. Alternatively, more complex models can be calibrated from hydrodynamical simulations and then interfaced with \texttt{oLIMpus}, either through tabulated data or via emulators. We leave this extension as future work}. 

\begin{figure}[ht!]
    \centering
    \includegraphics[width=\linewidth]{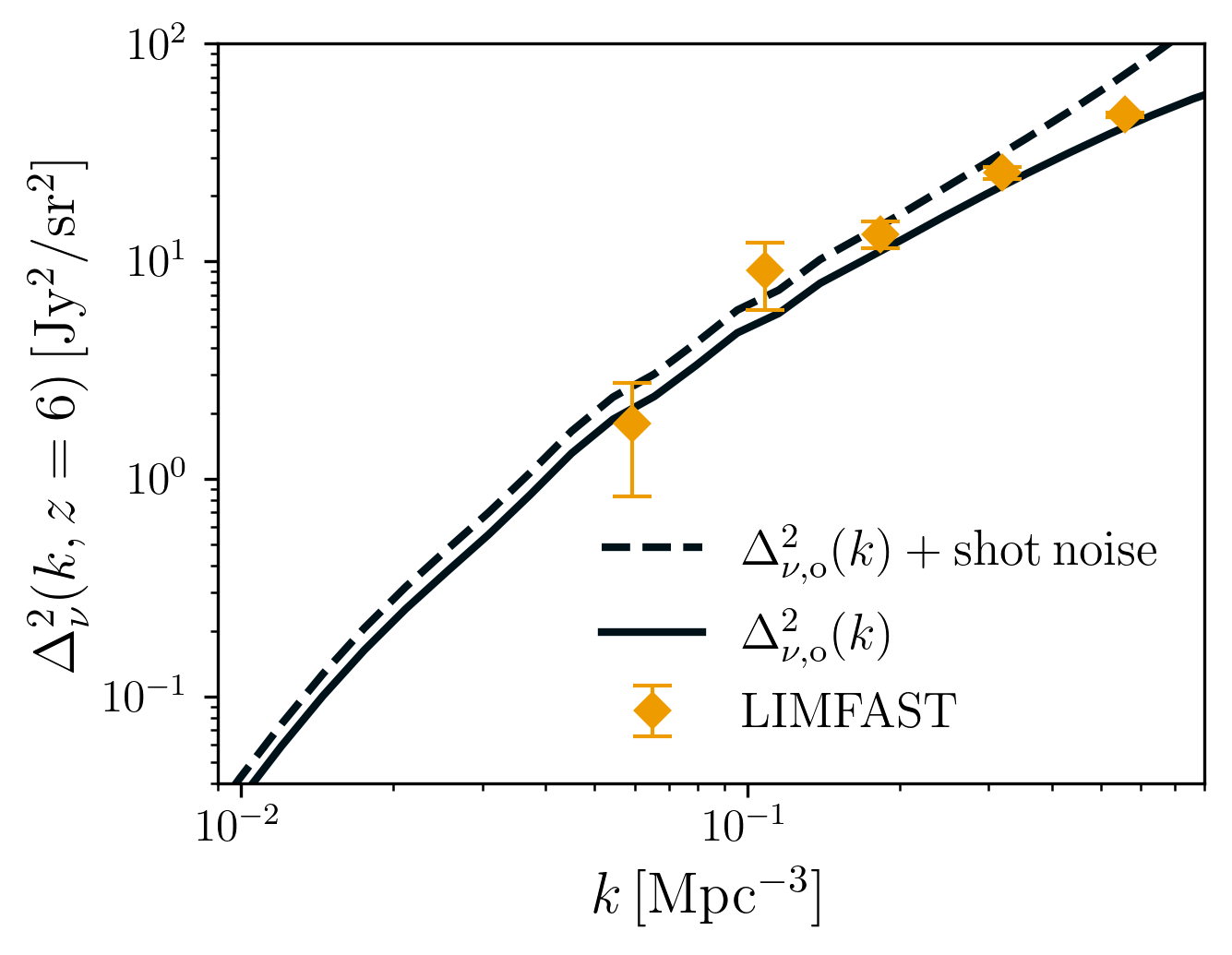}
    \caption{Power spectra computed using \texttt{oLIMpus}, with (black dashed) and without (black solid) shot noise  compared with the power spectrum measured on a \{$L_{\rm box},N_{\rm cell}$\}=\{150\,Mpc,150\} box produced with \texttt{LIMFAST}, without shot noise (yellow dots). The errobars indicate the $3\sigma$ mean standard error estimated through \texttt{powerbox}. The two codes show agreement on the scales accessible through the \texttt{LIMFAST} run, but the effective model of \texttt{oLIMpus} allows the user to quickly explore a wider range of scales. }
    \label{fig:compare_power_spectra}
\end{figure}

\begin{table*}[t!]
    \centering
    \renewcommand{\arraystretch}{1.5}
    \begin{tabular}{|P{2.35cm}|P{3.7cm}P{3.7cm}P{3.7cm}P{3.7cm}|}
    \hline
         Current release & \texttt{oLIMpus\footnoteref{olimpuslink}} & \texttt{lim}~\cite{Bernal:2019jdo} & \texttt{LIMFAST}~\cite{Mas-Ribas:2022jok,Sun:2022ucx,Sun:2024vhy}  & \texttt{21cmFASTv4}~\cite{Murray:2020trn,Mesinger_2011,Davies:2025wsa} \\
    \hline
    code type & analytical$+$mocks & analytical & semi-analytic simulation & semi-analytic simulation\\
    computational cost & light (analytical) or medium (box, lightcone) & light & {heavy} & heavy \\
    easy to change line/astro/cosmo & \cmark & \cmark & \xmark & \xmark \\
    \hline
    density evolution & linear using $D(z)$ & linear using $D(z)$ & non-linear 2LPT & non-linear 2LPT\\
    non-linearities in structures & through lognormal approximation & through halofit & cell-by-cell calculation & cell-by-cell calculation \\
    one-halo contribution & \xmark & \cmark & \xmark & \xmark \\
    \hline
    metallicity dependence & $\bcancel{\text{\cmark}}$ & \xmark & \cmark & \cmark\\
    feedbacks on star formation  & $\bcancel{\text{\cmark}}$ & \xmark & \cmark & \cmark\\
    reionization & \cmark (*) & \xmark & \cmark & \cmark\\
    \hline
    line luminosity & scaling relations & scaling relations & SED & post processing\\
    stochasticity & \cmark lognormal distribution & \cmark lognormal distribution & \xmark & \cmark sampling \\ 
    21-cm signal & \cmark & \xmark & \cmark & \cmark\\
    \hline
    shot noise & \cmark analytical & \cmark analytical & \xmark & \cmark from sampling \\
    redshift space distortions & analytical & analytical & velocity field & velocity field \\
    \hline
    observational properties & $\bcancel{\text{\cmark}}$(*) & \cmark & \makecell[t]{{\cmark external libraries}} & \makecell[t]{{\cmark external libraries}}\\
    \hline
    \end{tabular}
    \caption{Comparison between the characteristics of the current version of \texttt{oLIMpus} with the current releases of other relevant publicly available codes that can be used to estimate the LIM power spectrum during the EoR. This table is not definitive, since all the codes are actively used and can be modified and improved in the future. The $\bcancel{\text{\cmark}}$ marks indicate characteristics that are not yet included in the default setup of \texttt{oLIMpus}, but can be easily added to its formalism{, as discussed in the main text}. We marked with (*) the sections of \texttt{oLIMpus} or \texttt{Zeus21} that we are currently improving, which will be included in upcoming works~\cite{Sklansky:2025,Libanore:2025gte}. More detail and links to all the GitHub repositories can be found in the main text.}
    \label{tab:code_compare}
\end{table*}

Despite these differences, it is interesting to compare the analytical LIM power spectrum with the output of the simulation. We do so in Fig.~\ref{fig:compare_power_spectra}, where we show the LIM power spectrum computed with \texttt{oLIMpus} and the power spectrum measured using \texttt{powerbox} on a \texttt{LIMFAST} box. The scales that we can access through the \texttt{LIMFAST} run are limited compared with the analytical code, due to the higher computational cost of the simulation; we compromised by choosing a $1\,{\rm Mpc}$ resolution, comparable to the value of $R_0$ in the \texttt{oLIMpus} run. \texttt{LIMFAST} does not account for shot noise, since it does not resolve single halos, but it directly works using the density fields; for this reason, for comparison, we also plot the \texttt{oLIMpus} power spectrum without shot noise. Remarkably, the two codes show agreement in the range of scales that can be accessed from the \texttt{LIMFAST} run we performed; \texttt{oLIMpus} has the advantage of being able to probe at the same time large and small scales, in a short computational time.

Finally, we briefly comment on the possibility of producing LIM power spectra using \texttt{21cmFASTv4}. The main difference between this code and all the others we mentioned is the sampling procedure through which it is possible to account for discrete sources. This naturally introduces stochasticity in the galaxy-to-halo relation, and provides a numerical way to estimate the shot noise. Line luminosities in \texttt{21cmFASTv4} can be introduced by post-processing the gridded galaxy field produced during the simulation, consistently with the other radiation fields produced throughout the simulation. For the moment, the line post-processing is not implemented by default in \texttt{21cmFASTv4}, but it is left to the user.

To close this section, in table~\ref{tab:code_compare} we summarize the main characteristics of these codes compared with \texttt{oLIMpus}. {More than what we have discussed so far, we note that \texttt{lim} is currently the only code (among those considered here) that directly accounts for observational effects (e.g., instrumental noise). \texttt{LIMFAST} and \texttt{21cmFASTv4}, on the other hand, can be easily interfaced with external libraries such as \texttt{21cmSense}~\cite{Pober:2012zz,Pober:2013jna,Murray:2024the}, which provide a more complete and extensive treatment of the noise contribution. In future work, we plan to extend \texttt{oLIMpus} to include observational effects, both through the analytical approach adopted in \texttt{lim}, and by interfacing it with other codes to generate noise at the map level~\cite{Fronenberg_2024} or directly on the summary statistics.}


\begin{figure*}
    \centering
    \includegraphics[width=\linewidth]{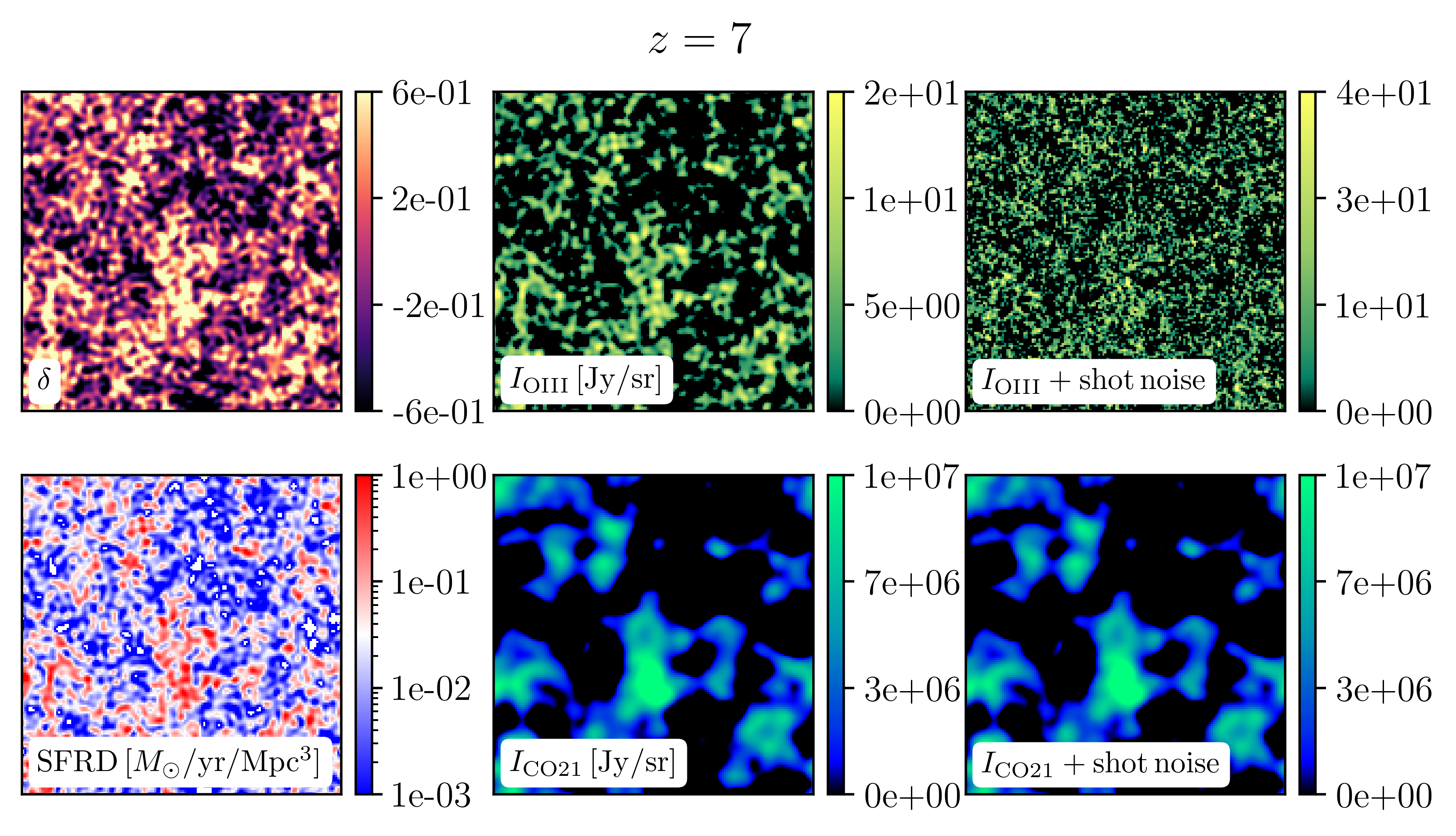}
    \caption{\texttt{oLIMpus} coeval boxes at $z=6$. On the left: density field $\delta$, produced passing $P_m(k,z)$ as input to \texttt{powerbox} (top); star formation rate density (SFRD), computed cell-by-cell through Eq.~\eqref{eq:rho_deltaR} (bottom). In the middle: OIII (top) and CO (bottom) intensities $\bar{I}_\nu$, obtained by passing $P_\nu(k,z)$ from Eq.~\eqref{eq:LIM_Pk} to \texttt{powerbox}; both these maps include RSD. For illustration, we considered $R=1\,{\rm Mpc}$ for OIII, $R=5\,{\rm Mpc}$ for CO. On the right: OIII (top) and CO(2-1) (bottom) intensities with shot noise, obtained adding a second box based on $P_{\rm shot}$ to the intensity. All boxes are produced using $L_{\rm box}=150\,$Mpc, $N_{\rm cell}=150$. 
    } 
    \label{fig:coeval_boxes}
\end{figure*}

\section{Map and lightcone making}\label{sec:maps}

In addition to producing LIM power spectra in $t\sim \mathcal{O}(\rm s)$, \texttt{oLIMpus} builds on the engine of {\tt Zeus21} and can thus generate coeval boxes of many relevant quantities:
\begin{itemize}[itemsep=1pt,topsep=5pt]
    \item the density $\delta(\vec{x},z)$, and smoothed versions of it $\delta_R(z)$;
    \item the star formation rate density $\dot{\rho}_*(\vec{x},z)$ and the luminosity density $\rho_L(\vec{x},z)$;
    \item the line intensity $\bar{I}_\nu(\vec{x},z)$, with and without shot noise and redshift space distortions, potentially smoothed over a certain $R$;
    \item the 21-cm brightness temperature and the ionization and bubble fields.
\end{itemize}
Each box has its own implementation, which we briefly describe in the next subsection; the boxes are produced at different $z$ independently. 

Figure~\ref{fig:coeval_boxes} shows an example of the boxes \texttt{oLIMpus} can produce: here, we show the density, the SFRD and the line intensity of OIII and CO(2-1), with and without shot noise. For illustration purposes, we produce the OIII box using $R=1\,$Mpc, and the CO box using $R=5\,{\rm Mpc}$ to mimic a lower-resolution observation. The correlation between fluctuations in the intensity and SFRD with the underlying density field is evident by eye. 

\begin{figure}
    \centering
    \includegraphics[width=\linewidth]{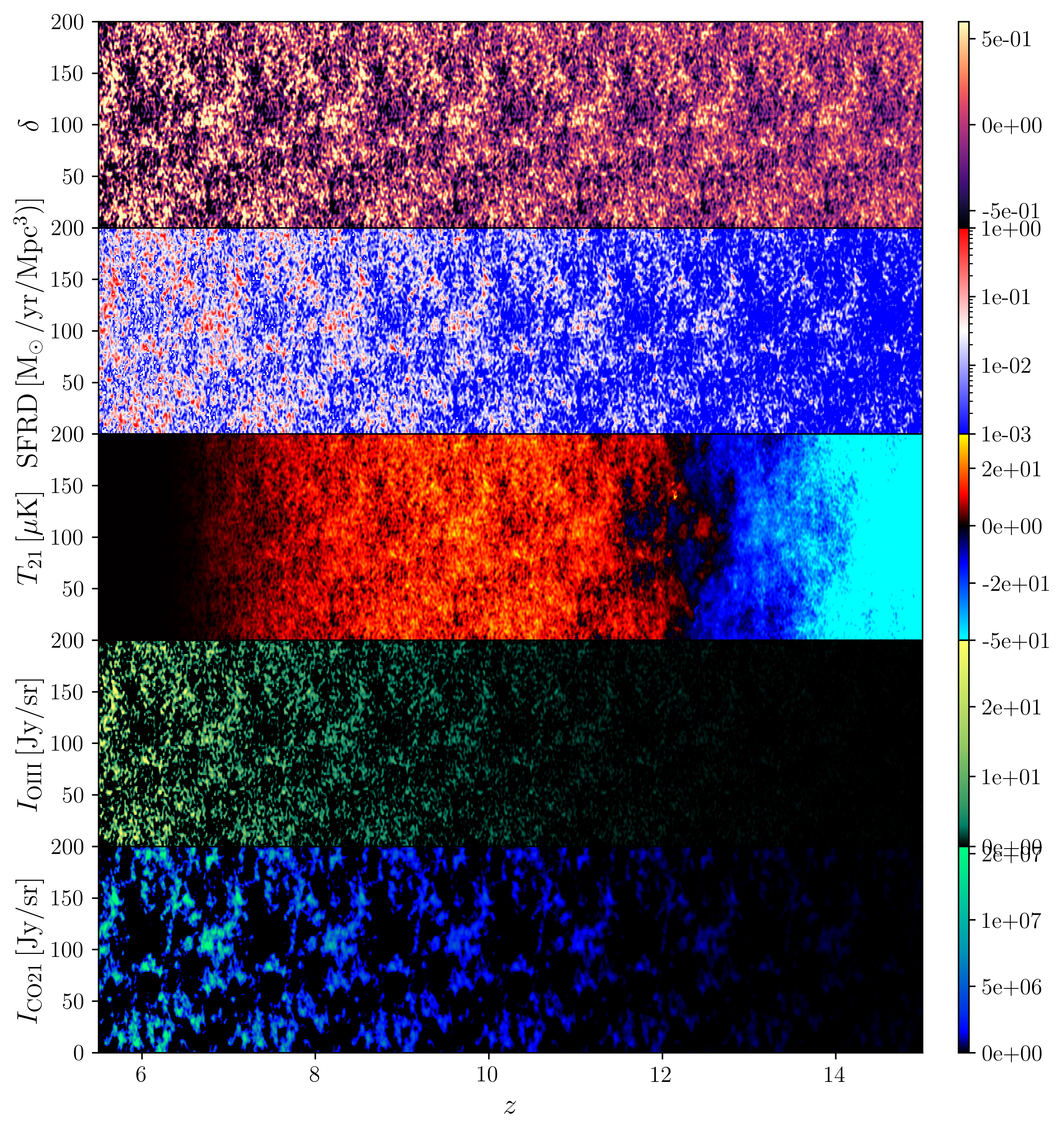}
    \caption{Illustrative example of the lightcones that can be produced using \texttt{oLIMpus}. From the top: density field, star formation rate density, 
    21-cm brightness temperature (only temperature, as in the original \texttt{Zeus21}; the bubble field will be included in the next release~\cite{Sklansky:2025}), OIII line intensity, CO(2-1) line intensity. In the case of CO, for illustration, we used $R_0=5\,{\rm Mpc}$.  
    } 
    \label{fig:lightcones}
\end{figure}

The code can also produce lightcones, as shown in Fig.~\ref{fig:lightcones}. To do so, it follows a similar procedure  to \texttt{21cmFAST}, that is, it produces a set of coeval boxes over a coarsely sampled redshift range, then it uses contiguous slices of either the same box or subsequent boxes to interpolate over a much finer array. If  the number of cells per side $N_{\rm cell}$ is too small, this procedure introduces spurious repetitions along the LoS.

\subsection{Coeval box implementation} 

We summarize how \texttt{oLIMpus} computes the different coeval boxes; we omit the description of the 21-cm and ionization fields, since these are not the main target of this paper but will be better explored in upcoming work~\cite{Libanore:2025gte}.

\subsubsection{Density field}

The density box $\delta(\vec{x},z)$ is produced using \texttt{powerbox}, which generates a Gaussian field with statistical properties described by $P_m(k,z)=D^2(z)P_m(k,z=0)$.

To smooth the density field to $\delta_R(\vec{x},z)$, we Fourier transform the $\delta(\vec{x},z)$ box,  multiply it by the squared top-hat window function $W^2(k,R)$ and then inverse-Fourier transform back to real space.

\subsubsection{Line intensity}

To produce the LIM intensity box $I_\nu(\vec{x},z)$, first of all we use \texttt{powerbox} to produce a Gaussian field with power spectrum $P_\nu(k,z)$ from either Eq.~\eqref{eq:pLIM_rsd} or~\eqref{eq:LIM_Pk}, depending on whether RSD are included or not.
Alternatively, \texttt{oLIMpus} can produce the $I_\nu(\vec{x},z)$ box by scaling the luminosity density box $\rho_L(\vec{x},z)$ computed cell-by-cell (see next section). This is the benchmark box we have been using throughout the paper to test our analytical model. In this case, RSD cannot be included. 

In both cases, we can then produce a second Gaussian field, with scale-independent power spectrum $P_{\rm shot}(z)$ from Eq.~\eqref{eq:shot}; this is generated using a different seed in \texttt{powerbox} to avoid introducing spurious correlations. 

Finally, we sum the intensity and shot noise field to obtain the full LIM coeval box. 
To conclude, the box is smoothed over the required radius $R$ using the same approach as in the density case.

\subsubsection{Star formation rate density and luminosity density}

As anticipated at the end of Sect.~\ref{sec:line_lum}, the SFRD and luminosity density boxes can be produced by computing cell-by-cell the density-weighted quantities. This is done first in Lagrangian space, then scaled to Eulerian space as in Eq.~\eqref{eq:rho_deltaR}. 
Shot noise can be introduced similarly to the previous section, while RSD cannot be included.

\section{Final remarks}\label{sec:conclusion}

We are entering an exciting era for line intensity mapping (LIM), with a growing number of instruments that are being commissioned or are coming online to survey line emission over large volumes. While pathfinder data and low-redshift results are already available, the upcoming years will see observations sensitive enough to finally access LIM during the epoch of reionization. Here, the details of star and galaxy formation are still unknown, and many open questions exist about the nature of the sources that ionized the intergalactic medium. 
In this context, it is becoming increasingly important to have tools at hand that can efficiently and consistently model the emission of multiple lines, while remaining capable of accurately describing the underlying physical processes.

With this in mind, in this paper we developed \texttt{oLIMpus}\footnoteref{olimpuslink}, an effective model to compute LIM auto- and cross- power spectra. \texttt{oLIMpus} builds on the pre-existing \texttt{Zeus21} code, which was developed in Refs.~\cite{Munoz:2023kkg,Cruz:2024fsv} to effectively model the 21-cm power spectrum. Both codes adopt a lognormal approximation for the density-weighted star formation rate density and line luminosity density, which are used to compute the smoothed two-point function fully analytically. 

In this paper, we improve the previous work in different aspects. First, we extend the computation to second-order lognormal approximations; this is required to accurately model the larger density fluctuations that arise at lower redshift and smaller smoothing radii. We developed the formalism to properly model the power spectra of star-forming lines, also including the shot noise and redshift space distortion contributions. The emission of OII, OIII, H$\alpha$, H$\beta$, CII, CO is parametrized through scaling relations calibrated on state-of-the-art simulations, with the possibility of accounting for lognormal stochasticity in the luminosity-to-halo mass relation. 

The \texttt{oLIMpus} code can compute the LIM auto- and cross- power spectra between star-forming lines and can produce coeval boxes and lightcones of their fluctuations across redshifts. Since it is built with \texttt{Zeus21} as a submodule, the code can also output 21-cm maps and power spectra. {Moreover, our framework can be extended to galaxy surveys by introducing a halo-occupation-distribution model (HOD) in Eq.~\eqref{eq:rho_Lag}. In this case, $f(M_h,z)=\langle N_g|M_h,z\rangle$ would represent the probability of finding $N_g$ galaxies inside a halo of mass $M_h$ at redshift $z$, and the result of the integral would be the galaxy number density. The subsequent part of the modeling should be updated accordingly and a detection threshold should be introduced.}

Below we highlight some prospects for the use of a code such as \texttt{oLIMpus}.

The low computational cost and high versatility make \texttt{oLIMpus} well-suited for data analysis, including MCMC searches over large parameter spaces. The code can be easily customized to include different cosmological, astrophysical, and line models. While it produces redshift-dependent power spectra of any line in a matter of seconds, it can also generate boxes and maps quickly and efficiently. These can be analyzed on a field level or with second- and higher-order summary statistics, or post-processed to include instrumental noise, foregrounds and line interlopers. We will explore these applications in future works.

At the same time, the presence of the luminosity-to-halo probability distribution function in the modeling offers the opportunity to explore more complex scenarios. Even if the current structure of \texttt{oLIMpus} does not support the use of stellar synthesis codes, as e.g.,~ is implemented in \texttt{LIMFAST}~\cite{Mas-Ribas:2022jok,Sun:2022ucx,Sun:2024vhy}, our formalism provides an elegant way to bridge results from more complete simulations to analytical relations, which serve to clearly and efficiently highlight inter-dependencies among relevant physical quantities. A compelling example can be found in the Balmer lines, H$\alpha$ and H$\beta$. Even if, in this first release, we modeled them through scaling relations, their luminosity is intrinsically related to the number of ionizing photons, which source them via recombinations of the hydrogen atoms both inside galaxies and in the intergalactic medium, see e.g.,~Ref.~\cite{1995ApJS...96....9L}. \texttt{oLIMpus} is well suited to investigate the relation between the emission of these lines and the ionization status of the intergalactic medium. We plan to improve the modeling of the Balmer lines and discuss these aspects in a dedicated work.
{Another compelling case is the Lyman-$\alpha$ line. Observationally, it provides valuable insights, as it is directly linked both to the star formation process, and to the evolution of IGM temperature and ionization. From the modeling perspective, however, Lyman-$\alpha$ is more challenging than other lines, since it requires connecting the intrinsic luminosity of star-forming regions to its propagation and effects in the IGM. Nevertheless, it is particularly interesting to explore, given that it will also be one of the main targets of SPHEREx, which will probe this line out to very high redshifts. We plan to address Lyman-$\alpha$ modeling in a future dedicated work. }

Finally, \texttt{oLIMpus} stands out as a powerful tool to consistently model and investigate cross-correlations. 

An immediate application, which we will soon present in a follow-up analysis~\cite{Libanore:2025gte}, is to study how the correlation between 21-cm and star-forming lines evolves between cosmic dawn and the epoch or reionization. At first, both signals have higher intensity in more dense regions, where more luminous sources are formed. With the onset of reionization and bubble formation, the two signals start to become anti-correlated. This switch from positive to negative correlation is expected no matter what the underlying cosmological or astrophysical models are. For this reason, cross-correlation is often advocated as a way to confirm signal detection on both 21-cm and other line-intensity maps, e.g.,~Refs.~\cite{Gong:2011mf,Visbal_2010,kovetz2017lineintensitymapping2017status}. Moreover, the scale where this switch occurs, the moment and duration of the transition are sensitive to variations in the parameters. Given the available 21-cm interferometers, and the large number of detectors targeting the other lines, this kind of analysis will soon offer unique insights on the physics of the epoch or reionization; for example, LOFAR~\cite{LOFAR:2013jil} and COMAP~\cite{COMAP:2021nrp} already envision such analysis in their near-future survey strategies.\footnote{P. Breysse, private communication.} 

Another interesting use for \texttt{oLIMpus} is its application to SPHEREx~\cite{SPHEREx:2014bgr}. The different frequency channels in its detector will collect emission from many lines across different redshifts. More than just using these maps separately, a consistent model for the line emissions will be crucial to study internal cross-correlations between channels. These can be used, for example, to improve noise treatment, reduce the contribution of interlopers and probe different phases in the IGM~\cite{Serra:2016jzs,Cheng:2024nfy,Roy:2023pei}. 

{Introducing an HOD model, \texttt{oLIMpus} could also be used to cross correlate the 21-cm and LIM signals with spectroscopic galaxy surveys, such as the ones that the Euclid Space Telescope~\cite{EUCLID:2011zbd} or the Nancy Grace Roman Space Telescope~\cite{Spergel:2015sza} will provide.}

With these perspectives, we strongly believe that a tool such as \texttt{oLIMpus} can be extremely valuable to the LIM community. While we will continue to improve its implementation and modeling, we have made the code publicly available, along with explanations and tutorials for its use and customization. We welcome readers to explore our GitHub repository\footnoteref{olimpuslink} and to contact us with any questions or suggestions.


\begin{acknowledgments}
We invite the reader to visit our GitHub repository\footnoteref{olimpuslink} and to explore the \texttt{oLIMpus} implementation; an informative tutorial is provided to guide through the first stages. 
We would like to thank Yonatan Sklansky, Emilie Thelie, Adam Lidz and Patrick Breysse for useful discussion. {We thank the anonymous referee for their valuable comments, many of which offer useful guidance for interesting follow-up work..}
SL is supported by an Azrieli International Postdoctoral Fellowship. EDK acknowledges joint support from the U.S.-Israel Bi-national Science Foundation (BSF, grant No.\,2022743) and the U.S. National Science Foundation (NSF, grant No.\,2307354), as well as support from the ISF-NSFC joint research program (grant No.\,3156/23). 
JBM acknowledges support from NSF Grants AST-2307354 and AST-2408637, and the CosmicAI institute AST-2421782.
SL thanks the University of Texas, Austin for hospitality during the last stages of this work; the visit was funded by the BGU/Philadelphia academic bridge grant, supported by the
Sutnick/Rosen Families Endowment Fund, Alton Sutnick/Helen Bershad BGU/Philadelphia Arts Collaboration Fund, The Alton Sutnick and Bettyruth Walter BGU/Philadelphia Collaboration Endowment Fund, The Mona Reidenberg Sutnick Memorial/Paula Bursztyn Goldberg Philadelphia Collaboration Endowment Fund, and the Sutnick/Zipkin BGU/Philadelphia Collaboration Expansion Fund.

\end{acknowledgments}

\appendix
\section{Emission line models in \texttt{oLIMpus}}\label{app:lines}

In this Appendix, we summarize the line-luminosity models $L(M_h)$ or $L(\dot{M}_*(M_h))$ which we included in \texttt{oLIMpus}. Our choice was to introduce two alternative models for the optical and UV rest-frame lines OII, OIII, H$\alpha$ and H$\beta$, one model for the infrared CII line and two models for the sub-mm lines CO(2-1) and CO(1-0). 

For each line, we decided to implement as default one or two models that are widely used in the literature, each calibrated on state-of-the-art simulations. On one side, this ensures that \texttt{oLIMpus} outputs can be easily compared with previous results. On the other, however, it also means that each line relies on slightly different assumptions and parameters, hence a one-to-one comparison between lines is not immediate. This may seem a bit confusing, but at the same time the code allows for easy customization through which the user can include any other model or even new lines. 

\subsection{Optical and UV lines}

Our baseline for the rest-frame optical and UV lines (OII, OIII, H$\alpha$, H$\beta$) is Ref.~\cite{yang2025newframeworkismemission}, in which the following scaling relation is calibrated on the Feedback in Realistic Environment simulation (FIRE~\cite{Ma_2018,Ma_2019,Ma_2020}) simulations:
\begin{equation}\label{eq:L}
    L(\dot{M}_*)\,[L_\odot] = \frac{2N\dot{M}_*}{(\dot{M}_*/{\rm SFR}_1)^{-\alpha_{\rm LIM}}+(\dot{M}_*/{\rm SFR}_1)^{\beta_{\rm LIM}}}, 
\end{equation}
where the values of $N$ and $\rm SFR_1$ depend on the line under analysis; we summarize their values in Tab~\ref{tab:L_pars}. The OIII luminosity estimated from Eq.~\eqref{eq:L} has been used to produce all the plots in the paper (except when differently specified).

\begin{table}[ht!]
    \centering
    \begin{tabular}{|c|cccc|}
        \hline
        & OIII & OII & H$\alpha$ & H$\beta$\\ 
        \hline
    $\lambda_{\rm rest}$\,[\AA] & 4960 & 3727 & 6563 & 4861 \\
    $\nu_{\rm rest}\,[{\rm GHz}]$ &  6.04$\times 10^{5}$ & 8.05$\times 10^{5}$ & 4.57$\times 10^{5}$ & 6.17$\times 10^{5}$ \\    
    \hline
    \hline
    $N$ & 2.75$\times 10^{7}$  & 2.14$\times 10^{6}$ &  4.54$\times 10^{7}$ & 1.61$\times 10^{7}$\\
    $\rm SFR_1$ &  1.24$\times 10^{2}$ & 5.91$\times 10^{1}$ &
3.81$\times 10^{1}$ & 1.74$\times 10^{1}$\\
    $\alpha_{\rm LIM}$ & 9.82$\times 10^{-2}$ & -2.43$\times 10^{-1}$ & 
 9.94$\times 10^{-3}$ & 7.98$\times 10^{-3}$\\
    
    $\beta_{\rm LIM}$ & 6.90$\times 10^{-1}$ & 2.50 & 5.25$\times 10^{-1}$ &  5.61$\times 10^{-1}$ \\
        \hline
        \hline
        $a$ & 7.84 & 7.08 & 8.08 & 7.62\\
        $m_a$ & 1.24 & 1.11 & 0.96 & 0.96 \\
        $m_b$ & 1.19 & 1.31 & 0.88 & 0.86 \\
        $m_c$ & 0.53 & 0.64 & 0.45 & 0.41\\
        $x_c$ & 0.66 & 0.54 & 0.96 & 0.96\\
        \hline
    \end{tabular}
    \caption{Fiducial parameters for Eq.~\eqref{eq:L} from  Ref.~\cite{yang2025newframeworkismemission} (mid panel) and for Eq.~\eqref{eq:L_THESAN} from Ref.~\cite{Kannan:2021ucy} (bottom panel). Despite OIII is a doublet, here we only quote one of the two lines (the one expected to be blended with H$\beta$ in SPHEREx observations~\cite{Gong_2017}). Parameters for OIII 5007\,\AA, as well as for far-infrared lines, are also provided in the references.}
    \label{tab:L_pars}
\end{table}

To study the variability of our results depending on the line model, we also implement a scaling relation calibrated on THESAN~\cite{THESAN_1,THESAN_2,THESAN_3}, a set of radiation-magneto-hydrodynamic simulations in the EoR. This scaling relation reads as~\cite{Kannan:2021ucy}
\begin{equation}\label{eq:L_THESAN}
    \log_{10} L\,[L_\odot] = \begin{cases}
        &(1)\quad a+m_a\log_{10}\dot{M}_*\\
        &(2)\quad a+(m_a-m_b)+m_b\log_{10}\dot{M}_*,\\
        &(3)\quad a+(m_a-m_b)+(m_b-m_c)x_c+ \\ 
        &\quad\quad+ m_c\log_{10}\dot{M}_*
    \end{cases}
\end{equation}
where $(1)$ holds when $\log_{10}\dot{M}_*<1$, (2) if $1\leq \log_{10}\dot{M}_*<x_c$ and (3) otherwise.
Also in this case, the parameters depend on the line under analysis. We quote the relevant values in Tab.~\ref{tab:L_pars}. 

\begin{figure*}[t!]
    \centering
   \includegraphics[width=.49\linewidth]{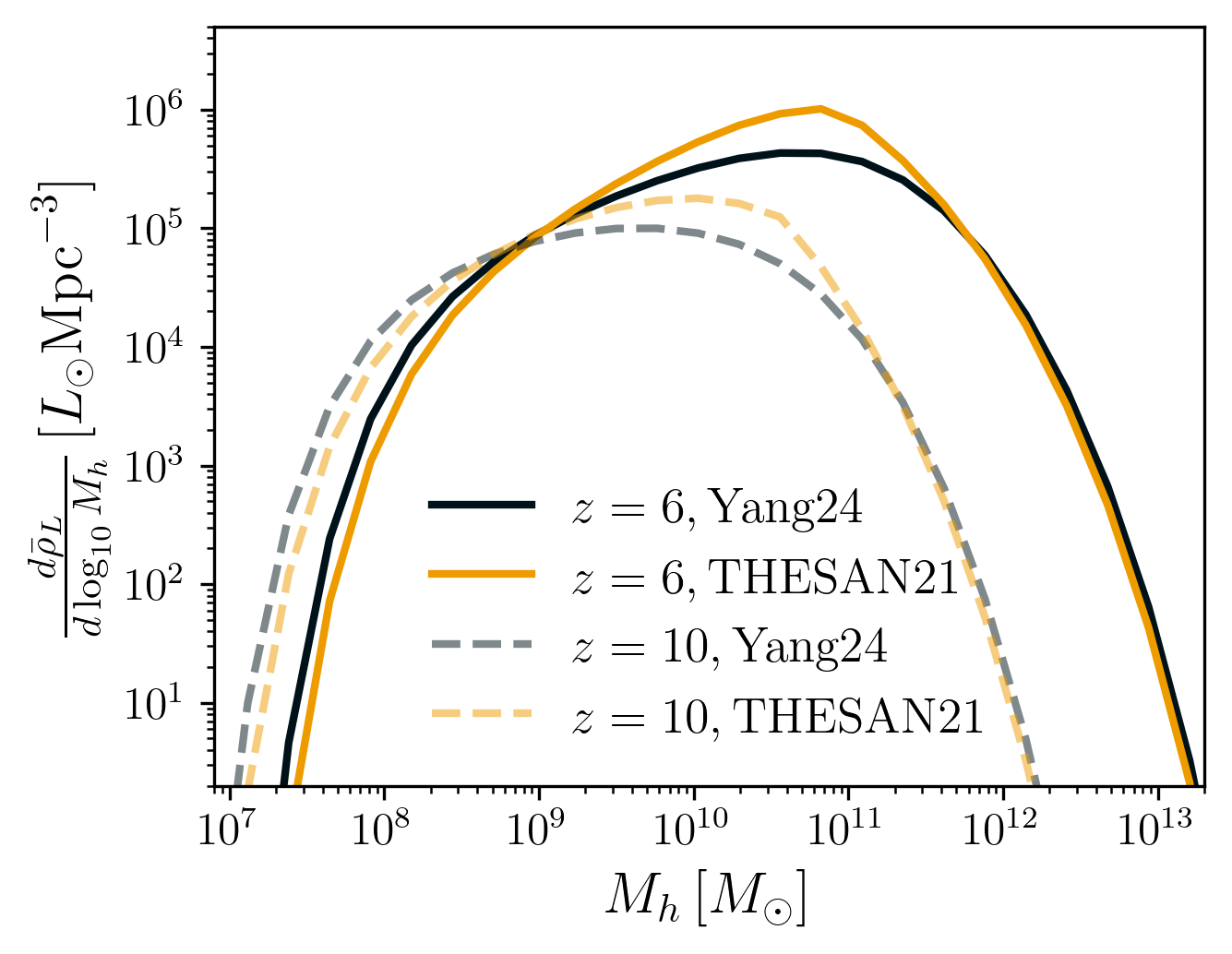}
   \includegraphics[width=.49\linewidth]{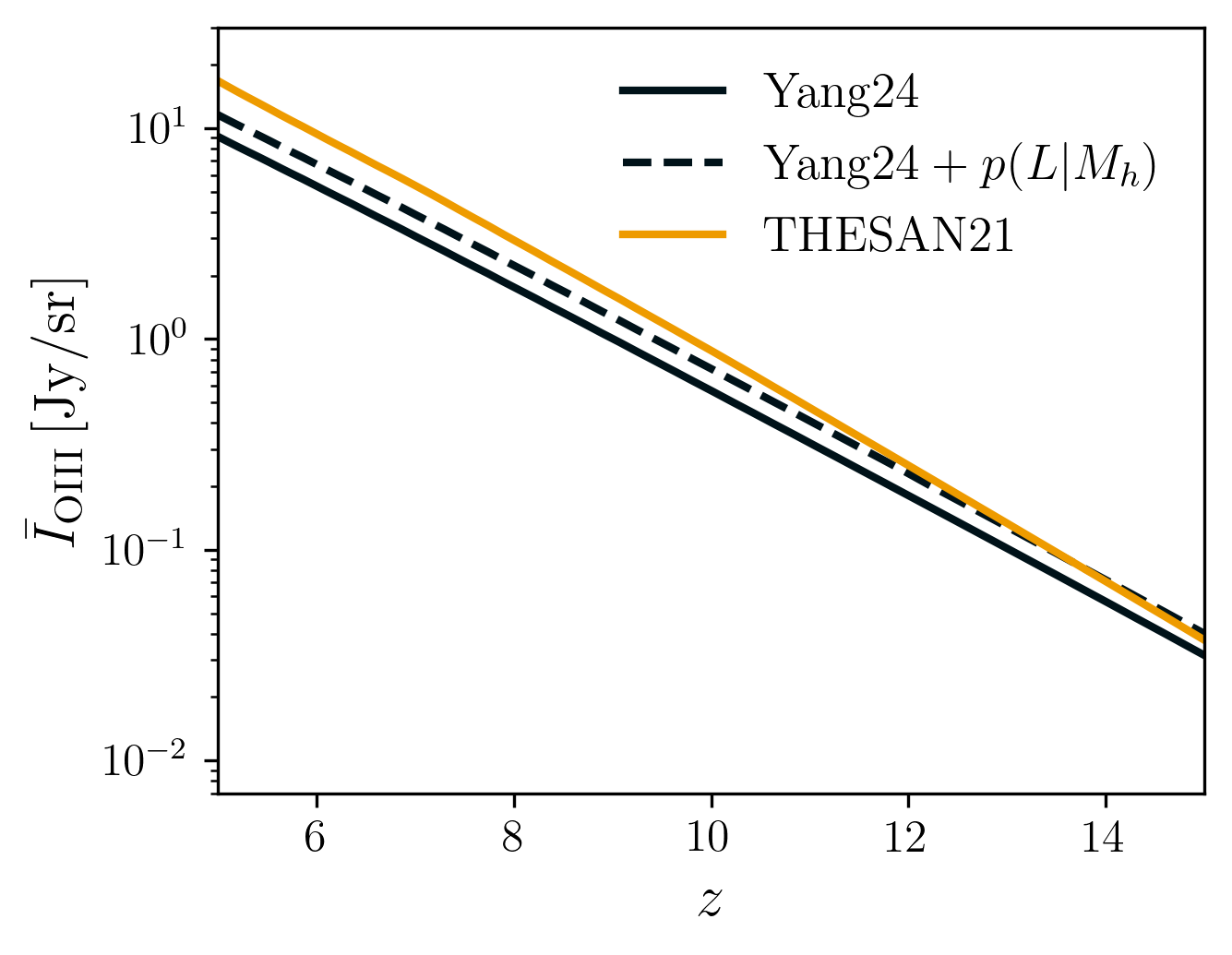}
    \caption{Left: Derivative of the OIII luminosity density with respect to $\log_{10}M_h$; this quantity shows how much each logarithmic bin of halo mass contributes to $\bar{\rho}_L$. Right: observed LIM intensity. In both panels, solid lines represent $z=6$, while the dashed $z=10$; black shows the model from Ref.~\cite{yang2025newframeworkismemission} (Yang24) while yellow from Ref.~\cite{Kannan:2021ucy} (THESAN21). }
    \label{fig:rho_L}
\end{figure*}

To understand how the different models characterize the line emission, in the left panel of Fig.~\ref{fig:rho_L} we show $d\bar{\rho}_L/d\log M_h$ for the OIII line, computed at $z=\{6,10\}$. This quantity characterizes the relative contribution of each logarithmic $M_h$ bin to the overall luminosity density. In both the models, this function mainly reflects the behavior of the underlying star formation rate (compare with Fig.~\ref{fig:SFR} in Appendix~\ref{app:SFR}), but some differences can be seen on the intermediate- and low-mass end. On the right panel, instead, we show how $\bar{I}_\nu(z)$  changes in the two models and when stochasticity from Eq.~\eqref{eq:stoch} is included (only for Ref.~\cite{yang2025newframeworkismemission} for clarity). For illustration purposes, we set $\sigma_L=0.7 = 0.3\,{\rm dex}$; this slightly increases the expected value of the observed intensity.  
Our choice of introducing the stochasticity in the $L(M_h)$ relation is different from, e.g.,~the approach in Ref.~\cite{Nikolic:2023wea}, in which a chain of integrals is used to introduce the PDFs $p(L'|\dot{M_*})$, $p(\dot{M_*}|M_*)$, $p({M_*}|M_h)$. Although we agree on the necessity of including stochastic contributions to describe the diversity of galaxy properties, we consider $p(L|M_h)$ a leaner choice, that nonetheless captures the key stochastic relationships among the most relevant quantities. 

\vspace*{-.3cm}
\subsection{Ionized Carbon (CII)}

The carbon fine-structure line CII at 158\,$\mu$m is emitted in the dense photo-dissociation regions in the outer layers of molecular clouds~\cite{Stacey:2010ps}. To model it, we rely on the model developed in Ref.~\cite{Lagache_2018}, 
\begin{equation}
    \log_{10}L = (a_{\rm SFR,0}+\alpha_{\rm SFR}z)\log_{10}\dot{M}_*+(\beta_{\rm SFR,0}+\beta_{\rm SFR}z), 
\end{equation}
where parameters are calibrated at $z<8$ combining the semi-analytic galaxy model from Ref.~\cite{Cousin_2015}, with the \texttt{CLOUDY} photoionization code~\cite{Ferland:2013dba}. 

However, we verified that the fiducial parameters from Ref.~\cite{Lagache_2018} return unphysical results when the luminosity model is extrapolated to high $z$. We therefore made the conservative choice of setting $\alpha_{\rm SFR}=\beta_{\rm SFR}=0$, and we set the parameters $\alpha_{\rm SFR,0}=0.7$ and $\beta_{\rm SFR,0}=3.4$ to their values at $z=10$ in the original model. 
More accurate modeling will be explored in the future.

\vspace*{-.3cm}
\subsection{Carbon Monoxide (CO)}

The CO molecule produces a series of rotational lines transitioning between the $J\to J-1$ levels, whose rest frame wavelengths are found as $\lambda_{J\to J-1}=2.6\,{\rm mm}/ J$. Different rotational lines can be observed in the sub-mm when originated from different redshifts; in particular, CO(2-1) is the target for EoR studies, see e.g.,~Ref.~\cite{COMAP:2021nrp}.

To model CO in \texttt{oLIMpus}, we rely primarily on the empirical model in Ref.~\cite{Yang:2021myt}, where
\begin{equation}
    L(M_h)= A\frac{2NM_h}{(M_h/M_p)^{-a}+(M_h/M_p)^b},
\end{equation}
where $A=1$, while $\{N,M_p,a,b\}$ are redshift-dependent quantities. Their values are calibrated for each line on the semi-analytic galaxy formation model of the Santa Cruz group~\cite{Somerville_1999}, coupled with the sub-mm line emission model framework of Ref.~\cite{Popping_2019}; in \texttt{oLIMpus} we included the CO(2-1) and CO(1-0) cases. 
In the current release of \texttt{oLIMpus}, we extend the validity of this model up to higher redshifts, despite its parameters were only calibrated up to $z=9$; we will improve the line modeling in future work. All CO(2-1) plots in the paper are produced using this model. 

Alternatively, we included the $L(\dot{M}_*(M_h))$ CO(2-1) model from Ref.~\cite{Li:2015gqa}, where 
\begin{equation}
    L(\dot{M}_*(M_h))=4.9\times 10^{-5}(10^{-\beta_{\rm LIM}}L_{\rm IR})^{1/\alpha_{\rm LIM}}
\end{equation}
where the infrared luminosity is estimated as $L_{\rm IR}=\dot{M}_*/10^{-10}\delta_{\rm MF}$. 
The values $\{\alpha_{\rm LIM},\beta_{\rm LIM},\delta_{\rm MF}\}=\{1.11,0.6,1\}$ have been calibrated in Refs.~\cite{Li:2015gqa,Keating:2020wlx} based on semi-analytic simulations~\cite{Obreschkow_2009}. 

\section{Star formation rate model}\label{app:SFR}

Throughout the main text and as default choice in \texttt{Zeus21} and \texttt{oLIMpus}, we use the SFR prescription from Ref.~\cite{Sabti:2021xvh}. Here, the star formation rate is computed as
\begin{equation}\label{eq:SFR}
    \dot{M}_*(M_h,z) = \frac{dM_h}{dt}(M_h,z)f_*(M_h,z)f_{\rm duty}(M_h,z),
\end{equation}
where the accretion rate of gas onto halos is~\cite{Schneider:2020xmf,Dekel:2013uaa}
\begin{equation}
    \frac{dM_h}{dt}(M_h,z)= M_h\alpha_{\rm acc}H(z)(1+z),
\end{equation}
with $\alpha_{\rm acc}=0.79$ from simulations. In Eq.~\eqref{eq:SFR}, the gas-to-stars conversion efficiency $f_*(M_h,z)={\dot{M}_*}/{\dot{M_h}}$ is
\begin{equation}\label{eq:fstar}
\begin{aligned}
    f_*(M_h,z) &= \frac{\Omega_b}{\Omega_m} \frac{2\epsilon_*}{(M_h/M_c)^{-\alpha_*} + (M_h/M_c)^{- \beta_*}}\leq 1,\\
    \epsilon_* &= \epsilon_{z_p}\times 10^{d\log_{10}\epsilon_*/dz}(z-z_p),
\end{aligned}
\end{equation} 
with $\alpha_*>0,\,\beta_*<0$. In the baseline case, we set the fiducial parameters to the same values as Ref.~\cite{Munoz:2023kkg},~$\{\epsilon_{z_p},z_p,d\log_{10}\epsilon_*/dz,\log_{10}M_c,\alpha_*,\beta_*\}=\{0.1,8,0,11.48,0.5,-0.5\}$.

Finally, the duty cycle $f_{\rm duty}(M_h,z)$ in Eq.~\eqref{eq:SFR} is defined as~\cite{Oh:2001ex} 
\begin{equation}
\begin{aligned}
    f_{\rm duty}(M_h,z) &= \exp\left[-\frac{M_{\rm atom}(z)}{M_h}\right],\\M_{\rm atom}(z) &= 3.3\times 10^7 \left(\frac{1.+z}{21}\right)^{-3/2}.
\end{aligned}
\end{equation}

\begin{figure}[t!]
    \centering
    \includegraphics[width=\linewidth]{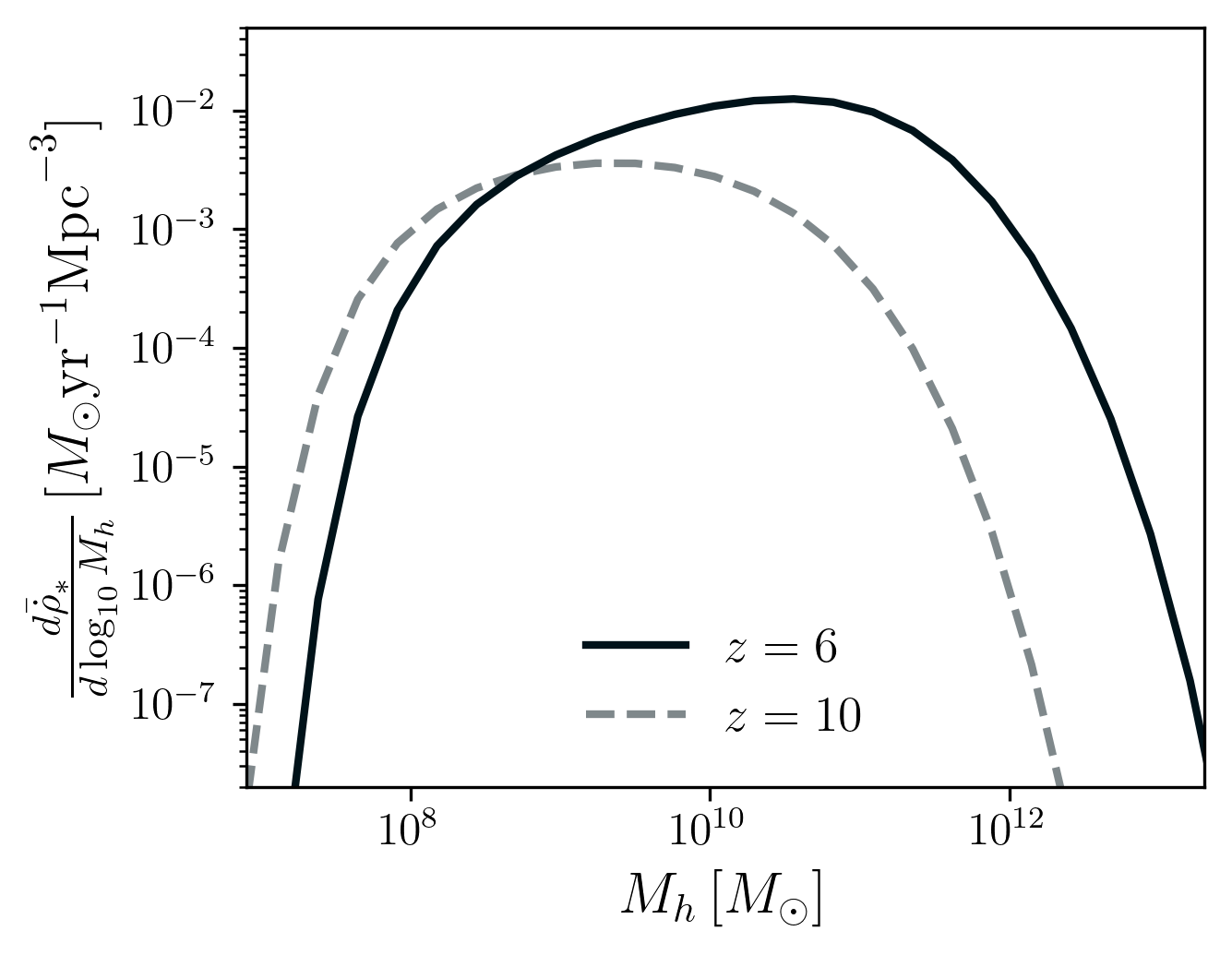}
    \caption{Derivative of the SFRD with respect to $\log_{10}M_h$. The comparison shows how much each logarithmic bin of $M_h$ contributes to the SFRD at $z=6$ (solid) and $z=10$ (dashed). }
    \label{fig:SFR}
\end{figure}

We use the star formation rate $\dot{M}_*(M_h)$ in Eq.~\eqref{eq:SFR} to compute the SFRD $\dot{\rho}_*$ (as well as the luminosity density $\rho_L$) through the equations in Sect.~\ref{sec:line_model}. To visualize how halos with different masses contribute to the SFRD, in Fig.~\ref{fig:SFR} we show $d\bar{\dot{\rho}}_*/d\log_{10}M_h$ at $z =\{6,10\}$. From the figure, it is clear that intermediate mass halos $M_h\sim 10^9M_\odot-10^{11}M_\odot$ dominate; moving to lower $z$, the relative contribution of more massive halos increases, thanks to the evolution of the HMF in the process of structure formation.

Moreover, for the moment \texttt{oLIMpus} and \texttt{Zeus21} rely on the deterministic prescription in Eq.~\eqref{eq:SFR} to compute the star formation rate. Including the contribution of stochasticity in the $\dot{M}_*-M_h$ relation, similarly to what we did for $L-M_h$ in Sect.~\ref{sec:line_lum}, would affect both the 21-cm and the LIM signal, changing the average values in the distributions described so far and adding fluctuations with astrophysical origin in their two-point functions (see e.g.,~Refs.~\cite{Nikolic:2023wea,Gagnon-Hartman:2025oxd,Davies:2025wsa}). We plan to explore these effects in \texttt{oLIMpus} with future work.

As a final remark, we note that, while \texttt{Zeus21} allows to include the contribution from population III stars (popIII) in mini-halos $M_h\simeq 10^7-10^8M_\odot$ to the evolution of the 21-cm signal~\cite{Cruz:2024fsv}, this choice is not yet implemented in \texttt{oLIMpus}. We consider this to be a good first-level approximation, since the contribution of popIII is mainly relevant at higher $z$ and lower metallicities than the ones required to produce the lines of interest for this work. Nevertheless, we plan to improve our code in the future, including an astrophysically motivated model for the contribution of popIII stars and mini-halos.

\section{Redshift Space Distortions}\label{app:RSD}

Distortions in the redshift space power spectrum arise due to perturbations in the observed positions of line emitters, caused by their peculiar velocities. 
To include them in our formalism, we follow the standard derivation of the Kaiser effect~\cite{Kaiser_1987}, as detailed e.g.,~in Refs.~\cite{peebles_1980,Dodelson:2003ft}. The generic $\rho$ we used in Sect.~\ref{sec:lognormal} to indicate either the SFRD or the line luminosity, is conserved under coordinate transformation between real ($\vec{x}$) and redshift space ($\vec{x}_s$). The two are related through the redshift $z$ and the peculiar velocity $\vec{v}$, since $x_s=z/H_0$ and $z=H_0x+\vec{v}\cdot\hat{x}$; using the Jacobian to describe the coordinate transformation and Taylor expanding it at linear order, we write
\begin{equation}
    \rho_{s}(\vec{x}_s) = J^{-1}\rho(\vec{x}) \simeq \left(1-\frac{\partial}{\partial x}\frac{\vec{v}\cdot\hat{x}}{H_0}\right)\rho(\vec{x}).
\end{equation}
In Sect.~\ref{sec:2pt}, we wrote the fluctuations in $\rho(\vec{x})$ in terms of $e^{\gamma_R\delta_R+\gamma_R^{\rm NL}\delta_R^2}$; we now use them to expand the computation to the redshift space, as
\begin{equation}\label{eq:xi_RS}
    \begin{aligned}
        \xi^{R_1R_2}_{s}&(r,z)=\langle \left(e^{\gamma_1\delta_{R_1}(\vec{x}_1)+\gamma_1^{\rm NL}\delta_{R_1}^2(\vec{x}_1)}-\frac{\partial}{\partial x}\frac{\vec{v}_1\cdot\hat{x}_1}{H_0}\right)\times\\
        &\quad\quad\quad\quad \times \left( e^{\gamma_2\delta_{R_2}(\vec{x}_2)+\gamma_2^{\rm NL}\delta^2_{R_2}(\vec{x}_2)}-\frac{\partial}{\partial x}\frac{\vec{v}_2\cdot \hat{x}_2}{H_0}\right)\rangle\\
        &=\langle e^{\gamma_1\delta_{R_1}(\vec{x}_1)+\gamma_1^{\rm NL}\delta_{R_1}^2(\vec{x}_1)}e^{\gamma_2\delta_{R_2}(\vec{x}_1)+\gamma_2^{\rm NL}\delta^2_{R_2}(\vec{x}_2)}\rangle+\\
        &\quad\quad -2\,\langle e^{\gamma_1\delta_{R_1}(\vec{x}_1)+\gamma_1^{\rm NL}\delta_{R_1}^2(\vec{x}_2)}\frac{\partial}{\partial x}\frac{\vec{v}_2\cdot\hat{x}_2}{H_0}\rangle+\\
        &\quad\quad +\langle \frac{\partial}{\partial x}\frac{\vec{v}_1\cdot\hat{x}_1}{H_0}\frac{\partial}{\partial x}\frac{\vec{v}_2\cdot\hat{x}_2}{H_0}\rangle.
    \end{aligned}
\end{equation}
The first line in the second equality is the real space two-point function, $\xi^{R_1R_2}(r,z)$. Viceversa, the last line represents the two-point function of the peculiar velocity. From the continuity equation we know that
\begin{equation}
    \tilde{v}(\vec{k})=i{f(z)H_0}\,\tilde{\delta}(\vec{k})\frac{\vec{k}}{k^2},
\end{equation}
where $f(z)=d\log D(z)/da=-(1+z)d\log D(z)/dz$ is the growth rate; hence, we can write 
\begin{equation}
\begin{aligned}
    \langle &\frac{\partial}{\partial x}\frac{\vec{v}_1\cdot\hat{x}_1}{H_0}\frac{\partial}{\partial x}\frac{\vec{v}_2\cdot\hat{x}_2}{H_0}\rangle \\
    &= \langle -{f^2}\frac{\partial}{\partial x}\left[\int \frac{d^3k}{(2\pi)^3}e^{i\vec{k}\cdot\vec{x}_1}\tilde{\delta}_1(\vec{k})\frac{\vec{k}\cdot \hat{x}_1}{k^2}\right]\times\\
    &\quad\quad\times\frac{\partial}{\partial x}\left[\int \frac{d^3k}{(2\pi)^3}e^{i\vec{k}\cdot\vec{x}_2}\tilde{\delta}_2(\vec{k})\frac{\vec{k}\cdot \hat{x}_2}{k^2}\right]\rangle\\
    &=\langle f^2 \int \frac{d^3k}{(2\pi)^3}\frac{k^2\mu^2}{k^2}e^{i\vec{k}\cdot\vec{x}_1}\tilde{\delta}_1(\vec{k})\int \frac{d^3k}{(2\pi)^3}\frac{\mu^2}{k^2}e^{i\vec{k}\cdot\vec{x}_2}\tilde{\delta}_2(\vec{k})\rangle\\
    &=(f\mu^2)^2\xi_m(r,z).
\end{aligned}
\end{equation}
In the second line we applied the derivatives to $e^{i\vec{k}\cdot\vec{x}_{1,2}}$, to get $k^2(\hat{k}\cdot \hat{x}_{1,2})\simeq k^2\mu_{1,2}$, where we approximated $\vec{x}$ with its line-of-sight (LoS) component. We defined $\mu$ as the cosine of the angle between $\vec{k}$ and the LoS; in the last line, we simply applied the Fourier transform to recover the matter two-point function. 

The two lines we have left in Eq.~\eqref{eq:xi_RS} represent the cross term between $\rho$ and the velocity field. Again we apply the Fourier transform on the velocity term, to get
\begin{equation}
\begin{aligned}
        \langle& e^{\gamma_1\delta_{R_1}(\vec{x}_1)+\gamma_1^{\rm NL}\delta_{R_1}^2(\vec{x}_1)}\frac{\partial}{\partial x}\frac{\vec{v}_2\cdot\hat{x}_2}{H_0}\rangle =\\ 
        &=-\langle e^{\gamma_1\delta_{R_1}(\vec{x}_1)+\gamma_1^{\rm NL}\delta_{R_1}^2(\vec{x}_1)}\int \frac{d^3k}{(2\pi)^3}f\mu^2 e^{i\vec{k}\cdot\vec{x}_2}\tilde{\delta}_2(\vec{k})\rangle\\
        &=-f\mu^2\xi^{R_1R_1}_{\nu m},
\end{aligned}
\end{equation}
where in the last line we re-arranged the integrals and the ensemble average to obtain the cross two-point function between $\rho$ and matter, 
\begin{align}
    &\langle e^{\gamma_1\delta_{R_1}(\vec{x}_1)+\gamma_1^{\rm NL}\delta_{R_1}^2(\vec{x}_2)}e^{\delta_0-\sigma_0^2/2}\rangle = e^{\frac{N^{R_1,0}}{D^{R_1,0}}-\log C^{R_1,0}}\\
    &N^{R_1,0}=\gamma_1\xi^{R_1,0}+\gamma_1^{\rm NL}(\xi^{R_1,0})^2+\frac{\gamma_1^2\sigma_1^2}{2}\\
    &D^{R_1,0}=1-2\gamma_1^{\rm NL}\sigma_1^2\\
    &C^{R_1,0}=\mathcal{N}_1\sqrt{D^{R_1,0}}.
\end{align}
For matter, $R_2=0$ and the $-\sigma_0^2/2$ in the exponent works as a normalization (instead of $\mathcal{N}_2$).
The expressions we obtained so far can be combined to get the power spectra in Eqs.~\eqref{eq:pLIM_rsd} and~\eqref{eq:pLIM_rsd_cross}. 

To be consistent with the choice in \texttt{Zeus21}, in \texttt{oLIMpus} we introduce three possible RSD options: 
\begin{itemize}[itemsep=1pt,topsep=5pt]
    \item $\mu=0$: power spectrum in real space;
    \item $\mu=0.6$: spherically averaged power spectrum, usually employed in simulations;
    \item $\mu=1$: only LoS modes, which in the case of 21-cm can be compared with the choice of observing outside the wedge to avoid foreground contamination;
\end{itemize} 
though we emphasize one can calculate the observable for many $\mu$ and integrate appropriately.

The derivation in this section holds on large scales, but breaks when the displacement in redshift space $\vec{v}\cdot\hat{x}/H_0$ becomes larger than the distance between pairs of observed points. On such small scales, elongated structures appear along the LoS, introducing non linearities in the treatment of RSD. These are the so-called Fingers-of-God (FoG); following Refs.~\cite{Peacock:1993xg,Ballinger:1996cd,Hamilton:1997zq}, we assume that the radial FoG distortion can be modeled as the convolution between the linear distortion and an incoherent velocity component that damps the power spectrum on large $k_\parallel=k\mu$ LoS scales. Therefore, we compute them by multiplying the linear RSD with the damping term $\mathcal{F}(k\mu\sigma_{\rm FoG})$, where $\sigma_{\rm FoG}\sim 7\,{\rm Mpc}$ depends on the halo dispersion velocity. Assuming an exponential distribution for the pairwise velocity~\cite{Davis:1982gc}, the FoG can be modeled as a Lorentzian factor in Fourier space, leading to
\begin{equation}
\begin{aligned}
    P^{\rm RSD+FoG}_{\nu}(k,z)&= {P_\nu^{\rm RSD}(k,z)}{\mathcal{F}(k\mu\sigma_{\rm FoG})}\\
    &=\frac{P_\nu^{\rm RSD}(k,z)}{[1+(k\mu\sigma_{\rm FoG})^2/2]^2},
\end{aligned}
\end{equation}
and similarly for the cross-power spectrum. 
We implemented the Lorentzian correction factor in \texttt{oLIMpus}, considering $\sigma_{\rm FoG}$ as an input free parameter. For the moment, our code allows the user to introduce the FoG effect only in the star-forming line power spectra, while we leave the study of their impact on the 21-cm power spectrum for future work. 

\bibliography{biblio.bib}

\end{document}